\documentclass[seceq]{ptptex}

\usepackage{wrapft} 
\usepackage[dvips]{graphicx}

\newcommand{\eqb}{\begin{equation}}
\newcommand{\eqe}{\end{equation}}
\newcommand{\eqbnon}{\begin{equation*}}
\newcommand{\eqenon}{\end{equation*}}

\newcommand{\eqab}{\begin{eqnarray}}
\newcommand{\eqae}{\end{eqnarray}}
\newcommand{\eqabnon}{\begin{eqnarray*}}
\newcommand{\eqaenon}{\end{eqnarray*}}

\newcommand{\seqb}{\begin{subequations}}
\newcommand{\seqe}{\end{subequations}}

\newcommand{\eref}[1]{\eqref{#1}} 

\newcommand{\defeq}{:=}
\newcommand{\defeqr}{=:}

\newcommand{\pd}[2]{\frac{\partial #1}{\partial #2}}

\newtheorem{ass}{Assumption}
\newtheorem{hypo}{Working Hypothesis}

\newtheorem{key}{Key Point}

\begin{document}

\markboth{Hiromi Saida}{To what extent is the entropy-area law universal ?} 

\title{To what extent is the entropy-area law universal ?}
\subtitle{Multi-horizon and multi-temperature spacetime may break the entropy-area law}

\author{Hiromi \textsc{Saida}\footnote{saida@daido-it.ac.jp}}

\inst{Department of Physics, Daido University, Minami-ku, Nagoya 457--8530, Japan}

\abst{
It seems to be a common understanding at present that, once event horizons are in thermal equilibrium, the entropy-area law holds inevitably. 
However no rigorous verification is given to such a very strong universality of the law in multi-horizon spacetimes. 
Then, based on thermodynamically consistent and rigorous discussion, this paper suggests an evidence of breakdown of entropy-area law for horizons in Schwarzschild-de~Sitter spacetime, in which the temperatures of the horizons are different. 
The outline is as follows: 
We construct carefully \emph{two thermal equilibrium systems} individually for black hole event horizon (BEH) and cosmological event horizon (CEH), for which the Euclidean action method is applicable. 
The integration constant (subtraction term) in Euclidean action is determined with referring to Schwarzschild and de~Sitter canonical ensembles. 
The free energies of the two thermal systems are functions of three independent state variables, and we find a similarity of our two thermal systems with the magnetized gas in laboratory, which gives us a physical understanding of the necessity of three independent state variables. 
Then, via the thermodynamic consistency with three independent state variables, the breakdown of entropy-area law for CEH is suggested. 
The validity of the law for BEH can not be judged, but we clarify the key issue for BEH's entropy. 
Finally we make comments which may suggest the breakdown of entropy-area law for BEH, and also propose two discussions; one of them is on the quantum statistics of underlying quantum gravity, and another is on the SdS black hole evaporation from the point of view of non-equilibrium thermodynamics.
}

\PTPindex{451, 454}

\maketitle

\section{Introduction}
\label{sec:intro}

Entropy-area law, which claims the equilibrium entropy of event horizon is equal to one-quarter of its spatial area in Planck units~\cite{ref:bht,ref:hr,ref:gsl}, is the equation of state of the event horizon in thermal equilibrium. 
This law has already been verified for spacetimes possessing a \emph{single} event horizon~\cite{ref:euclidean.sch,ref:euclidean.kn,ref:euclidean.sads,ref:euclidean.ds}. 
Then we may naively expect that the entropy-area law holds also for multi-horizon spacetimes, once every horizon is individually in thermal equilibrium. 
This expectation is equivalent to consider that the thermal equilibrium of each horizon is the necessary and sufficient condition to ensure the entropy-area law for each horizon. 
However this expectation has not been rigorously verified in multi-horizon spacetimes. 
(Comments on existing researches on Schwarzschild-de~Sitter spacetime will be given in fourth, fifth and sixth paragraphs in this section. 
Please wait for those paragraphs if the reader cares about existing researches on multi-horizon spacetimes.) 
At present, there remains the possibility that the thermal equilibrium may be simply the necessary condition of entropy-area law. 
If we find an example that some event horizon does not satisfy the entropy-area law even when it is in thermal equilibrium, then we recognize the thermal equilibrium as simply the necessary condition of the entropy-area law.

We can consider a situation in which the entropy-area law may break down in multi-horizon spacetime~\footnote{
All discussions in this paper are based on the ordinary general relativity. 
The other modified theories of gravity are not considered. 
Even if there is a breakdown of entropy-area law due to exotic fields of modified theory, such a breakdown in modified theory is out of the scope of this paper.
}. 
To explain it, it is necessary to \emph{distinguish} thermodynamic state of each horizon and that of the total system (multi-horizon spacetime) composed of several horizons. 
Even when every horizon in a multi-horizon spacetime is in an equilibrium state \emph{individually}, the total system composed of several horizons is never in any equilibrium state if the equilibrium state of one horizon is different from that of the other horizon. 
For example, if the temperatures of horizons in a multi-horizon spacetime are different from each other, then a net energy flow arises from a high temperature horizon to a low temperature one. 
Such multi-horizon spacetime can not be understood to be in any equilibrium state, since, exactly speaking, no energy flow arises in thermal equilibrium states. 
The thermal equilibrium of total system (multi-horizon spacetime) is realized if and only if the temperatures of all constituent horizons are equal. 
Therefore, when the temperatures of horizons are not equal, the multi-horizon spacetime should be understood as it is in a \emph{non-equilibrium} state. 
Here it should be noticed that, generally in non-equilibrium physics, once the system under consideration comes in a non-equilibrium state, the equation of state for non-equilibrium case takes different form in comparison with that for equilibrium case. 
Especially the non-equilibrium entropy deviates from the equilibrium entropy (when a non-equilibrium entropy is well defined). 
Indeed, although a quite general formulation of non-equilibrium thermodynamics remains unknown at present, the difference of non-equilibrium entropy from equilibrium one is already revealed for some restricted class of non-equilibrium systems~\cite{ref:noneq,ref:noneq.rad}.
Hence, for multi-horizon spacetimes composed of horizons of different temperatures, it seems to be reasonable to expect the breakdown of entropy-area law. 
However, since a ``non-equilibrium thermodynamics'' applicable to multi-horizon spacetime has not been constructed at present, we need to make use of ``equilibrium thermodynamics'' to investigate thermodynamic properties of multi-horizon spacetimes.

Motivated by the above consideration, this paper treats Schwarzschild-de~Sitter (SdS) spacetime as the representative of multi-horizon spacetimes. 
We construct \emph{two thermal equilibrium systems} in SdS spacetime; one of them is for black hole event horizon (BEH) and another is for cosmological event horizon (CEH). 
Note that, since the temperature of BEH is always higher than that of CEH in SdS spacetime~\cite{ref:temperature}, we need a good way to obtain thermal equilibrium systems of BEH and CEH. 
As will be explained in detail in Sec.\ref{sec:ass}, we will adopt the same way of constructing \emph{two} thermal equilibrium systems as Gibbons and Hawking have used in calculating the Hawking temperatures of BEH and CEH~\cite{ref:temperature}; it is to introduce a thin wall between BEH and CEH which reflects perfectly the Hawking radiation coming from the horizons. 
The region enclosed by the wall and BEH (CEH) settles down to a thermal equilibrium state, and we obtain two thermal equilibrium systems separated by the perfectly reflecting wall. 
Then we will examine the entropy-area law for the two thermal equilibrium systems \emph{individually}. 
(Although we are motivated by a non-equilibrium thermodynamic consideration in previous paragraph, the whole analysis in this paper is based on equilibrium thermodynamics and we discuss the \emph{two} equilibrium thermodynamics for BEH and CEH individually.) 
As will be explained in Sec.\ref{sec:ass}, our two thermal systems are treated in the canonical ensemble to obtain the free energies of BEH and CEH. 
Hence we will make use of the Euclidean action method which is regarded as one technique to obtain the partition function of canonical ensemble of quantum gravity~\cite{ref:euclidean}. 
(See Appendix~\ref{app:action} of this paper or of the previous paper~\cite{ref:euclidean.ds}.) 
Then we will find that the free energies are functions of three independent state variables. 
The existence of three independent state variables for SdS spacetime was not recognized in existing works on multi-horizon spacetimes~\cite{ref:temperature,ref:sds.special,ref:sds.existing.1,ref:sds.existing.2,ref:sds.effective}. 
But in this paper, thermodynamically rigorous analysis with three independent state variables will suggest a reasonable evidence of breakdown of entropy-area law for CEH. 
The validity of the law for BEH will not be judged, but we will clarify the key issue for BEH's entropy. 
These results imply that the thermal equilibrium of each horizon may not be the necessary and sufficient condition but simply the necessary condition of entropy-area law. 
The necessary and sufficient condition of the law may be implied via some existing works as noted below:

Let us note that some proposals for thermodynamics of BEH and CEH in SdS spacetime are already given to a case with some special matter fields and for an extreme case with magnetic/electric charge~\cite{ref:sds.special}. 
These examples are artificial to vanish the temperature difference of BEH and CEH, and show that the entropy-area law holds for SdS spacetime if the temperatures of BEH and CEH are equal. 
However, in this paper, we consider a more general case which is not extremal and does not include artificial matter fields. 
In all analyses in this paper, the temperatures of BEH and CEH remain different and the discussions in those examples~\cite{ref:sds.special} can not be applied. 
If we find the breakdown of entropy-area law for the case that horizons have different temperatures, then it is suggested that the necessary and sufficient condition of entropy-area law is the thermal equilibrium of the total system composed of several horizons in which the net energy flow among horizons disappears.

Next let us make comments on the case that horizons have different temperatures. 
The construction of SdS thermodynamics with leaving horizon temperatures different has already been tried in some existing works~\cite{ref:temperature,ref:sds.existing.1,ref:sds.existing.2,ref:sds.effective}. 
Those works assume some geometrical conserved quantities to be state variables of SdS spacetime, and derive the so-called mass formula which is simply a geometrical relation and looks similar to the first law of black hole thermodynamics. 
However, the \emph{thermodynamic consistency} has not been confirmed in those works. 
Here ``thermodynamic consistency'' means that the state variables satisfy not only the four laws of thermodynamics but also the appropriate differential relations; for example, the differential of free energy $F_o$ with respect to temperature $T_o$ is equivalent to the minus of entropy, $S_0 \equiv - \partial F_o/\partial T_o$~\footnote{
There are many other similar differential relations in thermodynamics. 
Those relations are the ones required in the ``thermodynamic consistency'', and necessary to understand thermodynamic properties of the system under consideration; e.g. phase transition, thermal and mechanical stabilities, and equations of states.
}.
If one asserts some theoretical framework to be a ``thermodynamics'', the framework must satisfy the thermodynamic consistency. 
Therefore, exactly speaking, it remains unclear whether those existing works~\cite{ref:temperature,ref:sds.existing.1,ref:sds.existing.2,ref:sds.effective} are appropriate as ``thermodynamics''~\footnote{
Those existing works~\cite{ref:temperature,ref:sds.existing.1,ref:sds.existing.2,ref:sds.effective} preserve/assume the entropy-area law without confirming thermodynamic consistency. Hence, if the breakdown of entropy-area law is concluded via the thermodynamic consistency, those existing works can not be regarded as ``thermodynamic'' theory.
}. 
On the other hand, it seems to be preferable that the number of assumptions for thermodynamic formulations of BEH and CEH is as small as possible. 
In order to introduce the minimal set of assumptions which preserves thermodynamic consistency, we will refer to Schwarzschild thermodynamics formulated by York~\cite{ref:euclidean.sch} (and also refer to de~Sitter thermodynamics~\cite{ref:euclidean.ds} which is also based on York~\cite{ref:euclidean.sch}).

Furthermore, motivated by the dS/CFT correspondence conjecture~\cite{ref:ds/cft}, some existing works~\cite{ref:sds.existing.2} focus their attention on the future and past null infinities in SdS spacetime (see Fig.\ref{fig:1} shown in Sec.\ref{sec:ass}, $I^{\pm}$ is the null infinities). 
Those infinities may be appropriate to discuss some geometrical quantities. 
However, as implied by the causal structure of SdS spacetime, the future null infinity seems to be inappropriate to discuss thermodynamic properties of BEH and CEH, because any observer near future null infinity (not near the future temporal infinity $i^+$) can not ``access'' BEH~\footnote{
The observer going towards the future temporal infinity $i^+$ in Fig.\ref{fig:1} can ``access'' BEH, since the BEH becomes a boundary of the causally connected region of that observer (the region~I in Fig.\ref{fig:1}).
}.

Hence, contrary to the existing works, the analysis in this paper is based on the following two points:
\begin{itemize}
\item 
As will be explained precisely in next section, we focus our attention on the region enclosed by BEH and CEH (not on null infinity) in SdS spacetime as the object of thermodynamic interests. 
\item 
We have a high regard to the ``thermodynamic consistency'' preserved by the minimal set of assumptions without referring to some geometrical conserved quantities and dS/CFT correspondence. 
\end{itemize}
Then, as the result of these two points, a suggestion of breakdown of entropy-area law will be obtained.

Here let us emphasize that, in next section, we exhibit explicitly the assumptions on which our discussion is based. 
We think readers can judge the approval or disapproval to every part of our assumptions and analyses. 
Therefore, even if some part of our discussion and analysis is not acceptable for some reader, we hope this paper can propose one possible issue about the universality of entropy-area law.

This paper is organized as follows: 
Sec.\ref{sec:ass} introduces the minimal set of assumptions in which, with referring to Schwarzschild and de~Sitter canonical ensembles~\cite{ref:euclidean.sch,ref:euclidean.ds}, the two thermal equilibrium systems for BEH and CEH are constructed, and the special role of the cosmological constant is also pointed out. 
Sec.\ref{sec:action} is for the calculation of Euclidean actions of the two thermal equilibrium systems. 
Sec.\ref{sec:beh} discusses thermodynamics for BEH and examines the entropy-area law for BEH. 
In that section, the validity of the law for BEH can not be judged, but the key issue for BEH's entropy is clarified. 
Sec.\ref{sec:ceh} proposes a reasonable evidence of the breakdown of entropy-area law for CEH. 
Finally Sec.\ref{sec:sd} is for summary and discussions, in which we make some physical comments which may suggest the breakdown of entropy-area law for BEH without rigorous verification, and propose two discussions; one of them is on the quantum statistics of underlying quantum gravity, and another is on the SdS black hole evaporation process from the point of view of non-equilibrium thermodynamics. 
And four appendices support the main text of this paper: 
Appendix~\ref{app:action} and~\ref{app:key} summarize, respectively, the Euclidean action method and the essence of York's Schwarzschild thermodynamic, Appendix~\ref{app:formulas} exhibits useful differential formulas for calculations of thermodynamic state variables, and Appendix~\ref{app:nariai} analyzes the Nariai limit (extremal limit) of our SdS thermodynamics. 
Throughout this paper we use the Planck units, $c = G = \hbar = k_B = 1$.

\section{Minimal set of assumptions}
\label{sec:ass}

\subsection{Preliminary}
\label{sec:ass.preliminary}

Let us dare to start this section with the discussion given already in de~Sitter canonical ensemble~\cite{ref:euclidean.ds}, because this discussion is conceptually essential for thermodynamic consistency.

The aim of this section becomes clear by considering the relation between thermodynamics and statistical mechanics~\cite{ref:sm,ref:ll}. 
In statistical mechanics, the partition function can not be expressed as a ``function of state variables'' unless the appropriate state variables, on which the partition function depends, are specified \emph{a priori}. 
To understand this, consider for example an ordinary gas in a spherical container of radius $L$, in which the number of constituent particles is $N$, the mass of one particle is $m$ and the mean velocity of particles is $v$. 
The ordinary statistical mechanics, without the help of thermodynamics, yields the partition function $Z_{\rm gas} = Z_{\rm gas}(L, N, m, v)$ as simply a function of ``parameters'', $L$, $N$, $m$ and $v$. 
Statistical mechanics, solely, can not determine what combinations of those parameters behave as state variables. 
To determine it, the first law of thermodynamics is necessary. 
(Note that the notion of \emph{heat} in the first law is established by purely the argument in thermodynamics, not in statistical mechanics.) 
Comparing the differential of partition function with the first law results in the identification of partition function with the free energy divided by temperature. 
Then, since the free energy of ordinary gases is a function of the temperature and volume due to the ``thermodynamic'' argument, the partition function $Z_{\rm gas}(L, N, m, v)$ should be rearranged to be a function of temperature and volume $Z_{\rm gas}(V, T)$, where $V = (4 \pi/3) L^3$ and $T = m v^2$ for ideal gases due to the law of equipartition of energy~\footnote{
When the number of particles $N$ changes by, for example, a chemical reaction and an exchange of particles with environment, $N$ is also the state variable on which the free energy depends.
}. (The dependence on $N$ is, for example, $Z_{\rm gas} \propto N$ for ideal gases.)

Here note that the reason why the temperature and volume are regarded as the state variables of the gas is that they are consistent with the four laws of ``thermodynamics'' and have the appropriate properties as state variable. 
The appropriate properties are that the state variables are macroscopically measurable, the state variables are classified into two categories, \emph{intensive} variables and \emph{extensive} variables, and the extensive variables are additive. 
Those properties of state variables are specified by purely the argument in thermodynamics, not in statistical mechanics. 
Therefore, from the above, it is recognized that statistical mechanics can not yield the partition function as a ``function of appropriate state variables'' without the help of thermodynamics which specifies the appropriate state variables for the partition function.

Turn our discussion to the Euclidean action method for curved spacetimes. 
Since the Euclidean action method is the technique to obtain the ``partition function'' of the spacetime under consideration (see Appendix~\ref{app:action}), it is necessary to specify the state variables before calculating the Euclidean action. 
In this section, referring to Schwarzschild and de~Sitter canonical ensembles~\cite{ref:euclidean.sch,ref:euclidean.ds}, we introduce the minimal set of assumptions for SdS thermodynamics, which specify the appropriate state variables for the partition function. 
Also, the special role of cosmological constant is clarified, which is already found in previous works~\cite{ref:euclidean.ds,ref:sds.effective}. 
The calculation of Euclidean action is carried out not in this section but in next section.

\subsection{SdS spacetime}

Before introducing the minimal set of assumptions, let us summarize the Lorentzian SdS spacetime in order to prepare some quantities used in the following discussions.

\begin{figure}[t]
 \begin{center}
 \includegraphics[height=35mm]{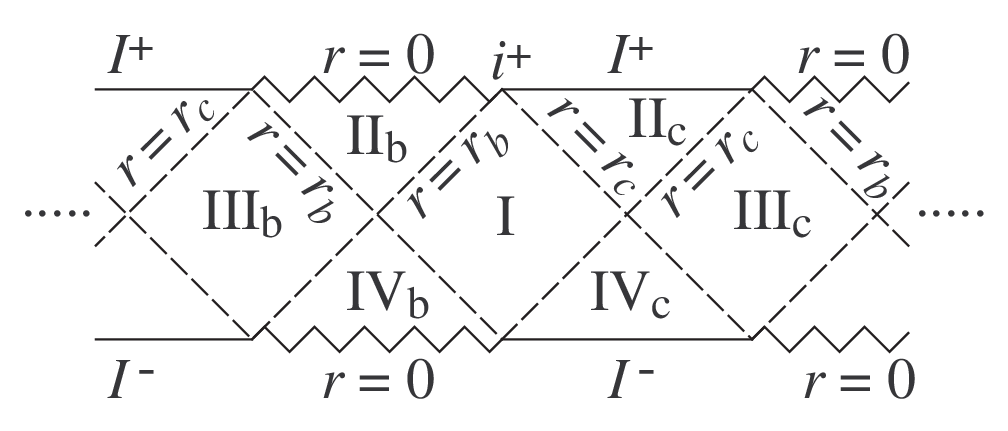}
 \end{center}
\caption{Penrose diagram of SdS spacetime. $I^{\pm}$ are the future/past null infinity, $i^+$ is the future temporal infinity. BEH is at $r=r_b$, and CEH at $r=r_c$. Spacetime singularity is at $r=0$. Static chart covers the region~I. Semi-global black hole chart covers the regions I, II$_b$, III$_b$ and IV$_b$. Semi-global cosmological chart covers the regions I, II$_c$, III$_c$ and IV$_c$. The maximal extension is obtained by connecting the two semi-global charts alternately.}
\label{fig:1}
\end{figure}

The metric of SdS spacetime in the \emph{static chart} is
\eqb
 ds^2 = -f(r)\,dt^2 + \frac{dr^2}{f(r)} + r^2\,d\Omega^2 \, ,
\label{eq:ass.SdS.static}
\eqe
where $d\Omega^2 = d\theta^2 + \sin^2\theta\,d\varphi^2$ is the line element on the unit two-sphere, and
\eqb
\label{eq:ass.SdS.f.H}
 f(r) \defeq 1 - \frac{2\,M}{r} - H^2\,r^2 \quad,\quad 3\,H^2 \defeq \Lambda \, ,
\eqe
where $M$ is the mass parameter of black hole and $\Lambda$ is the cosmological constant. 
The Penrose diagram of SdS spacetime is shown in Fig.\ref{fig:1}, and the static chart covers the region~I.

An algebraic equation $f(r)=0$ has one negative root and two positive roots. 
The smaller and larger positive roots are, respectively, the radius of BEH $r_b$ and that of CEH $r_c$. 
The notion of CEH is observer dependent and the CEH at $r_c$ is associated with the observer going towards the temporal future infinity $i^+$ in region~I~\cite{ref:temperature}. 
The equation $f(r) = 0$ is rearranged to $4\,\tilde{r}^3 -3\,\tilde{r} + \sqrt{27}\,M\,H = 0$, where $\tilde{r} \defeq \sqrt{3}\,H\,r/2$. 
Then via a formula, $\sin\theta = -4\,\sin^3(\theta/3) + 3\,\sin(\theta/3)$, we get
\eqb
\label{eq:ass.r}
 r_b = \frac{2}{\sqrt{3}\,H}\,\sin\left(\frac{\alpha}{3}\right) \quad,\quad
 r_c = \frac{2}{\sqrt{3}\,H}\,\sin\left(\frac{\alpha + 2 \pi}{3}\right) \, ,
\eqe
where $\alpha$ is defined by, $\sin\alpha \defeq \sqrt{27}\,M\,H$. 
The existence condition of BEH and CEH is $0 < \sqrt{27}\,M\,H < 1$. 
This is equivalent to, $0 < \alpha < \pi/2$, which means
\eqb
 2 M < r_b < 3 M < \frac{1}{\sqrt{3} H} < r_c < \frac{1}{H} \, .
\label{eq:ass.range.rb.rc}
\eqe
This denotes that $r_b$ is larger than the Schwarzschild radius $2 M$ and $r_c$ is smaller than the de~Sitter's CEH radius $H^{-1}$.

SdS spacetime has a timelike Killing vector $\xi \defeq N\, \partial_t$, where $N$ is a normalization constant~\cite{ref:sg}. 
This $\xi$ becomes null at BEH and CEH. 
This means those horizons are the Killing horizons of $\xi$. 
The surface gravity of BEH $\kappa_b$ and that of CEH $\kappa_c$ are defined by the equations, 
$\nabla_{\xi}\,\xi^{\mu}\,|_{r=r_b}
  = \kappa_b\,\xi^{\mu}\,|_{r=r_b}$ , 
$\nabla_{\xi}\,\xi^{\mu}\,|_{r=r_c}
  = \kappa_c\,\xi^{\mu}\,|_{r=r_c}$. 
The surface gravity depends on $N$. 
Throughout this paper we take the normalization $N = 1$ for BEH, and $N = -1$ for CEH to make $\kappa_c$ positive~\footnote{Even if $N = 1$ for CEH, we can keep consistency of our analysis by changing the signature of $\kappa_c$ appropriately.}. 
Then the surface gravities become equal to the absolute value $\left|(1/2)\,df(r)/dr \right|$ at each Killing horizon,
\eqb
\label{eq:ass.kappa}
\begin{split}
 \kappa_b &= \frac{H^2}{2 r_b}\,\left( r_c - r_b \right)\,\left( 2\,r_b + r_c \right)
           = \frac{1}{2\,r_b}\,\left( 1 - 3\,H^2\,r_b^2 \right) \, , \\
 \kappa_c &= \frac{H^2}{2 r_c}\,\left( r_c - r_b \right)\,\left( r_b + 2\,r_c \right)
           = \frac{1}{2\,r_c}\,\left( 3\,H^2\,r_c^2 - 1 \right) \, ,
\end{split}
\eqe
where Eq.\eref{eq:ass.r} is used in the second equality in each equation. 
From the inequality $r_b < r_c$ in Eq.\eref{eq:ass.range.rb.rc}, we get
\eqb
 \kappa_b > \kappa_c \, .
\label{eq:ass.kappa.rel}
\eqe
This implies the Hawking temperature of BEH is higher than that of CEH, which will be verified by Eqs.\eref{eq:beh.Tb} and~\eref{eq:ceh.Tc}.

For later use, let us show some differentials,
\seqb
\eqab
 \pd{\,r_b}{M} = \frac{2}{1 - 3\,H^2\,r_b^2} \quad&,&\quad
 \pd{\,r_b}{H} = -\frac{r_b}{H} + \frac{M}{H}\,\pd{\,r_b}{M}
\label{eq:ass.diff.rb} \, , \\
 \pd{\,r_c}{M} = - \frac{2}{3\,H^2\,r_c^2 - 1} \quad&,&\quad
 \pd{\,r_c}{H} = -\frac{r_c}{H} + \frac{M}{H}\,\pd{\,r_c}{M}
\label{eq:ass.diff.rc} \, ,
\eqae
and
\eqab
 \pd{\,\kappa_b}{M} = -\frac{1}{r_b^2}\,\frac{1 + 3\,H^2\,r_b^2}{1 - 3\,H^2\,r_b^2}
 \quad&,&\quad
 \pd{\,\kappa_b}{H}
   = \frac{1 - 3\,H^2\,r_b^2}{2\,H\,r_b}
    + \frac{M}{H}\,\pd{\,\kappa_b}{M}
\label{eq:ass.diff.kappab} \, , \\
 \pd{\,\kappa_c}{M} = -\frac{1}{r_b^2}\,\frac{3\,H^2\,r_c^2 + 1}{3\,H^2\,r_c^2 - 1}
 \quad&,&\quad
 \pd{\,\kappa_c}{H}
   = \frac{3\,H^2\,r_c^2 - 1}{2\,H\,r_c}
    + \frac{M}{H}\,\pd{\,\kappa_c}{M}
\label{eq:ass.diff.kappac} \, ,
\eqae
\seqe
where we used a formula, $\cos\theta = 4\,\cos^3(\theta/3) - 3\,\cos(\theta/3)$, and the differentials, $\partial_M \alpha = \sqrt{27} H/\cos\alpha$, and $\partial_H \alpha = \sqrt{27} M/\cos\alpha$, obtained from the definition of $\alpha$, $\sin\alpha \defeq \sqrt{27} M H$.

The metric in \emph{semi-global black hole chart} is given by the coordinate transformation from $(t,r,\theta,\varphi)$ to $(\eta_b,\chi_b,\theta,\varphi)$:
\eqb
 \eta_b - \chi_b \defeq - e^{-\kappa_b (t - r^{\ast})} \quad,\quad
 \eta_b + \chi_b \defeq e^{\kappa_b (t + r^{\ast})} \, ,
\label{eq:ass.trans.beh}
\eqe
where $dr^{\ast} \defeq dr/f(r)$ which means
\eqb
 2\,r^{\ast} =  \ln\left| \frac{r}{r_b} - 1 \right|^{1/\kappa_b}
              - \ln\left| 1 - \frac{r}{r_c} \right|^{1/\kappa_b}
              + \ln\left| \frac{r}{r_b + r_c} + 1 \right|^{1/\kappa_c-1/\kappa_b} \, .
\label{eq:ass.rstar}
\eqe
We get by this transformation,
\eqb
 ds^2 = \Upsilon_b(r)\,\left[\, - d\eta_b^2 + d\chi_b^2 \,\right] + r^2\,d\Omega^2 \, ,
\eqe
where
\eqb
 \Upsilon_b(r) \defeq
 \frac{2 M}{\kappa_b^2\,r}\,\left( 1 - \frac{r}{r_c} \right)^{1+\kappa_b/\kappa_c}\,
 \left(\frac{r}{r_b + r_c} + 1 \right)^{2 - \kappa_b/\kappa_c} \, .
\label{eq:ass.upsilon.beh}
\eqe
The transformation~\eref{eq:ass.trans.beh} implies the range of coordinates, $-\chi_b < \eta_b < \chi_b$ and $0 < \chi_b$, which covers the region~I in Fig.\ref{fig:1}. 
By extending to the range, $-\infty < \eta_b < \infty$ and $-\infty < \chi_b < \infty$, the semi-global black hole chart covers the regions I, II$_b$, III$_b$ and IV$_b$ in Fig.\ref{fig:1}. 
In these regions we find $\Upsilon_b > 0$ since $r < r_c$.

The metric in \emph{semi-global cosmological chart} is given by the coordinate transformation from $(t,r,\theta,\varphi)$ to $(\eta_c,\chi_c,\theta,\varphi)$:
\eqb
 \eta_c - \chi_c \defeq e^{\kappa_c (t - r^{\ast})} \quad,\quad
 \eta_c + \chi_c \defeq -e^{-\kappa_c (t + r^{\ast})} \, ,
\label{eq:ass.trans.ceh}
\eqe
where $r^{\ast}$ is given in Eq.\eref{eq:ass.rstar}. 
By this transformation we get
\eqb
 ds^2 = \Upsilon_c(r)\,\left[\, - d\eta_c^2 + d\chi_c^2 \,\right] + r^2\,d\Omega^2 \, ,
\eqe
where
\eqb
 \Upsilon_c(r) \defeq
 \frac{2 M}{\kappa_c^2\,r}\,\left( \frac{r}{r_b} - 1 \right)^{1+\kappa_c/\kappa_b}\,
 \left(\frac{r}{r_b + r_c} + 1 \right)^{2 - \kappa_c/\kappa_b} \, .
\label{eq:ass.upsilon.ceh}
\eqe
The transformation~\eref{eq:ass.trans.ceh} implies the range of coordinates, $\chi_c < \eta_c < -\chi_c$ and $\chi_c < 0$, which covers the region~I in Fig.\ref{fig:1}. 
By extending to the range, $-\infty < \eta_c < \infty$ and $-\infty < \chi_c < \infty$, the semi-global cosmological chart covers the regions I, II$_c$, III$_c$ and IV$_c$ in Fig.\ref{fig:1}. 
In these regions we find $\Upsilon_c > 0$ since $r_b < r$.

The maximally extended SdS spacetime is obtained by connecting the two semi-global charts alternately as shown in Fig.\ref{fig:1}.

\subsection{Minimal set of assumptions and working hypothesis}

As mentioned in Sec.\ref{sec:ass.preliminary}, we introduce the minimal set of assumptions with referring to Schwarzschild thermodynamics formulated by York~\cite{ref:euclidean.sch}. 
Those assumptions should construct thermal systems for SdS thermodynamics and give us enough state variables which appear as independent variables in the free energy of our thermal system. 
There are three key points in Schwarzschild thermodynamics from which we can learn about the way to ensure the ``thermodynamic consistency'' in SdS thermodynamics. 
Those key points are the same from which the basic assumptions of de~Sitter thermodynamics are introduced in previous paper~\cite{ref:euclidean.ds}, and we summarize those three key points in Appendix~\ref{app:key} of this paper.

Here we must comment on Anti-de~Sitter (AdS) black holes~\cite{ref:euclidean.sads}.
AdS black hole thermodynamics has a conceptual difference from the other black hole thermodynamics.
The difference appears, for example, in the definition of temperature. 
While the temperatures in asymptotic flat black hole and de~Sitter thermodynamics include the \emph{Tolman factor}~\cite{ref:euclidean.sch,ref:euclidean.kn,ref:euclidean.ds,ref:tolman}, the temperature assigned to AdS black hole~\cite{ref:euclidean.sads} does not include the Tolman factor, where the Tolman factor~\cite{ref:tolman} expresses the gravitational redshift affecting the Hawking radiation propagating from horizon to observer (see for example Eq.\eref{eq:beh.Tb} in Sec.\ref{sec:beh.1} and the key point~3 of Schwarzschild thermodynamics in Appendix~\ref{app:key}). 
The temperature in AdS black hole thermodynamics can not be measured by a thermometer of the physical observer who are outside the black hole. 
This implies that the state variables in AdS black hole thermodynamics are defined not by the observer outside the black hole, but defined just on the black hole event horizon on which no physical observer can rest.
In this paper we do not refer to AdS black hole thermodynamics, since it seems to be preferable to expect that state variables are defined according to a physical observer. 
Hence we refer to Schwarzschild thermodynamics, which is based on a physical observer (See Appendix~\ref{app:key}).

\subsubsection{Zeroth law and independent variables}
\label{sec:ass.assumption1}

\begin{figure}[t]
 \begin{center}
 \includegraphics[height=35mm]{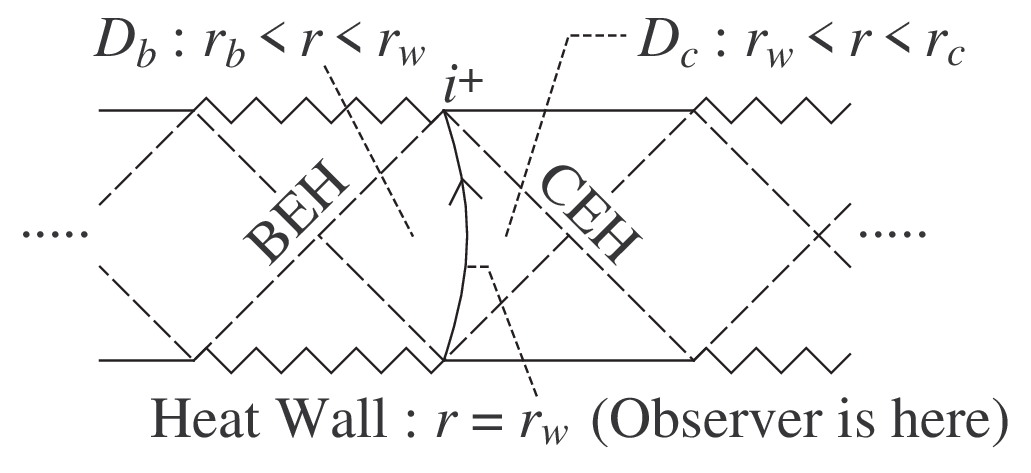}
 \end{center}
\caption{Our thermal equilibrium systems for BEH ($D_b$) and CEH ($D_c$). State variables of them are defined at the heat wall.}
\label{fig:2}
\end{figure}

We will construct two thermal equilibrium systems for BEH and CEH, but place only one observer who can measure the state variables of BEH and CEH. 
Such observer is in the region~I, $r_b < r < r_c$ (see Fig.\ref{fig:1}). 
However, as mentioned at Eq.\eref{eq:ass.kappa.rel}, Hawking temperature of BEH is higher than that of CEH. 
This temperature difference implies that, when the region~I constitutes one thermodynamic system, the thermodynamic state of region~I is in a non-equilibrium state. 
Therefore, by dividing the region~I into two regions, we construct \emph{two} thermal equilibrium systems for BEH and CEH individually which are measured by the same observer. 
To do so, we adopt the following assumption as the zeroth law: 
\begin{ass}[Zeroth law] 
Two thermal ``equilibrium'' systems for BEH and CEH in SdS spacetime are constructed by the following three steps:
\begin{enumerate}
\item
Place a spherically symmetric thin wall at $r = r_w$ in the region~I, $r_b < r_w < r_c$, as shown in Fig.\ref{fig:2}. 
This wall has negligible mass, and reflects perfectly Hawking radiation coming from each horizon. 
We call this wall the ``heat wall'' hereafter. 
The BEH side of heat wall is regarded as a ``heat bath'' of Hawking temperature of BEH due to the reflected Hawking radiation. 
Also the CEH side of heat wall is regarded as a heat bath of Hawking temperature of CEH.
\item
The region $D_b$ enclosed by BEH and heat wall, $r_b < r < r_w$, is filled with Hawking radiation emitted by BEH and reflected by heat wall, and forms a thermal equilibrium system for BEH. 
Similarly the region $D_c$ enclosed by CEH and heat wall, $r_w < r < r_c$, forms a thermal equilibrium system for CEH. 
Hence we have ``two'' thermal equilibrium systems separated by the heat wall. 
In the statistical mechanical sense, these two thermal equilibrium systems are described by the canonical ensemble, since those systems have a contact with the heat wall.
\item
Set our observer at the heat wall. 
When the observer is at the BEH side of heat wall, the observer can measure the state variables of thermal equilibrium system for BEH. 
The same is true of CEH. 
Then the state variables of two thermal equilibrium systems are defined by the quantities measured by the observer at heat wall.
\end{enumerate}
\end{ass}
Note that the two thermal equilibrium systems constructed in this assumption have already been used by Gibbons and Hawking~\cite{ref:temperature} to calculate Hawking temperatures of BEH and CEH. 
Also the above step~3, which gives a criterion of defining state variables, has already been adopted in the consistent thermodynamics of single-horizon spacetimes~\cite{ref:euclidean.sch,ref:euclidean.kn,ref:euclidean.ds}. 
We can regard this assumption as a simple extension of the key point~1 of Schwarzschild thermodynamics shown in Appendix~\ref{app:key}.

It is expected that state variables of the thermal equilibrium system for BEH depend on BEH radius $r_b$ and/or BEH surface gravity $\kappa_b$. 
Similarly, state variables of CEH depend on $r_c$ and/or $\kappa_c$. 
These imply that the state variables of BEH and CEH depend on $M$ and $\Lambda$, since the horizon radii and surface gravities depend on $M$ and $\Lambda$ via Eqs.\eref{eq:ass.r} and~\eref{eq:ass.kappa}. 
Furthermore, by the step~3 in assumption~1, there should be $r_w$-dependence in state variables of BEH and CEH, since the observer is at $r = r_w$. 
Therefore the state variables of BEH and CEH depend on three parameters $M$, $\Lambda$ and $r_w$.

The existence of three parameters $(M,\Lambda,r_w)$ may imply that the CEH is regarded as a source of external gravitational field which affects the thermodynamic state of BEH. 
Also, BEH is a source of external gravitational field affecting the thermodynamic state of CEH. 
Here it is instructive to compare \emph{qualitatively} our two thermal equilibrium systems of horizon with a magnetized gas. 
The gas consists of molecules possessing a magnetic moment, and its thermodynamic state is affected by an external magnetic field. 
The qualitative correspondence between the magnetized gas and our thermal equilibrium systems of horizon is described as follows; 
the gas corresponds to the system $D_b$~($D_c$), and the external magnetic field corresponds to the external gravitational field produced by CEH~(BEH). 
Then, what we should emphasize is the following fact of the magnetized gas: 
When the gas is enclosed in a container of volume $V_{\rm gas}$ and an external magnetic field $\vec{H}_{\rm ex}$ is acting on the gas, the free energy $F_{\rm gas}$ of the gas is expressed as a function of three independent state variables (see for example \S52, 59 and~60 in Landau and Lifshitz~\cite{ref:ll}),
\eqb
\label{eq:ass.Fgas}
 F_{\rm gas} = F_{\rm gas}(T_{\rm gas} , V_{\rm gas} , \vec{H}_{\rm ex}) \, ,
\eqe
where $T_{\rm gas}$ is the temperature of the gas. 
According to this fact of the magnetized gas, it is reasonable for our two thermal equilibrium systems of horizons to require that the free energies are also functions of three independent state variables. 
For the BEH, the free energy $F_b$ is
\eqb
\label{eq:ass.Fb}
 F_b = F_b(T_b , A_b , X_b) \, ,
\eqe
where $T_b$ is the temperature of BEH, $A_b$ is the state variable of system size, and $X_b$ is the state variable which represents the effect of CEH's gravity on the BEH. 
And for the CEH, the free energy $F_c$ is
\eqb
\label{eq:ass.Fc}
 F_c = F_c(T_c , A_c , X_c) \, ,
\eqe
where $T_c$ is the temperature of CEH, $A_c$ is the state variable of system size and $X_c$ is the state variable which represents the effect of BEH's gravity on the CEH. 
Indeed, it will be proven in Secs.~\ref{sec:beh.1} and~\ref{sec:ceh.1} that the thermodynamic consistency never hold unless the free energies are functions of three independent variables as shown in Eqs.\eref{eq:ass.Fb} and~\eref{eq:ass.Fc}.

Now we recognize that, because free energies are functions of three independent variables (as will be verified in Secs.~\ref{sec:beh.1} and~\ref{sec:ceh.1}), the following working hypothesis is needed: 
\begin{hypo}[Three independent variables] 
To ensure the thermodynamic consistency of our thermal equilibrium systems constructed in assumption~1, we have to regard the cosmological constant $\Lambda$ as an independent working variable. 
Then the three quantities $(M,\Lambda,r_w)$ are regarded as independent variables, and consequently the free energies $F_b$ and $F_c$ of our thermal equilibrium systems are functions of three independent state variables as shown in Eqs.\eref{eq:ass.Fb} and~\eref{eq:ass.Fc}. 
On the other hand, when we regard the non-variable $\Lambda$ as the physical situation, it is obtained by the ``constant $\Lambda$ process'' in the consistent thermodynamics for BEH and CEH which are constructed with regarding $\Lambda$ as a working variable.
\end{hypo}

This working hypothesis will be verified in Secs.~\ref{sec:beh.1} and~\ref{sec:ceh.1}, and we can not preserve thermodynamic consistency without regarding $\Lambda$ as an independent working variable. 
This implies that, as already commented in Sec.V of previous paper~\cite{ref:euclidean.ds}, \emph{the thermal equilibrium states of event horizon with positive $\Lambda$ may construct the ``generalized'' thermodynamics in which $\Lambda$ behaves as a working variable and the physical process is described by the constant $\Lambda$ process.}

\subsubsection{Scaling law and system size}

In thermodynamics of ordinary laboratory systems, all state variables are classified into two categories, \emph{extensive} state variables and \emph{intensive} ones. 
The criterion of this classification is the scaling behavior of state variables under the scaling of system size. 
However, as explained by the key point~2 of Schwarzschild thermodynamics shown in Appendix~\ref{app:key}, the state variables in thermodynamics of single-horizon spacetimes~\cite{ref:euclidean.sch,ref:euclidean.kn,ref:euclidean.ds} have its own peculiar scaling behavior classified into three categories, and the state variable of system size is not a volume but the surface area of heat bath.
Although the scaling behavior differs from that in thermodynamics of ordinary laboratory systems, the peculiar scaling behavior in thermodynamics of single-horizon spacetimes retains the thermodynamic consistency as explained in the key point~3 in Appendix~\ref{app:key}. 
Then, we assume that the key point~2 of Schwarzschild thermodynamics is simply extended to our two thermal equilibrium systems constructed in the assumption~1:
\begin{ass}[Scaling law and system size]
All state variables of our thermal equilibrium systems are classified into three categories; extensive variables, intensive variables and thermodynamic functions, and satisfy the following scaling law: 
When the length size $L$ (e.g. horizon radius) is scaled as $L \to \lambda\,L$ with an arbitrary scaling rate $\lambda\,(>0)$, then the extensive variables $X$ (e.g. system size) and intensive variables $Y$ (e.g. temperature) are scaled respectively as $X \to \lambda^2\,X$ and $Y \to \lambda^{-1}\,Y$, while the thermodynamic functions $\Phi$ (e.g. free energy) are scaled as $\Phi \to \lambda\,\Phi$. 
This implies that the thermodynamic system size of our thermal equilibrium systems should have the areal dimension, since the system size is extensive in thermodynamic argument. 
Then we assume that the surface area of heat wall, $A \defeq 4 \pi r_w^2$, behaves as the consistent extensive variable of system size for our thermal equilibrium systems for BEH and CEH. 
This denotes to set $A_b = A_c = A$ in Eqs.\eref{eq:ass.Fb} and~\eref{eq:ass.Fc}.
\end{ass}

Accepting this assumption, the length size scaling in our thermal equilibrium systems for BEH and CEH should be specified. 
Here recall that, due to the working hypothesis~1, the fundamental independent variables in our thermal equilibrium systems are $M$, $\Lambda$ and $r_w$. 
Therefore the fundamental length size scaling for our thermal equilibrium systems is composed of the following three scalings;
\eqb
\label{eq:ass.scaling}
 M \to \lambda\,M \quad,\quad
 H \to \frac{1}{\lambda}\,H \quad,\quad
 r_w \to \lambda\,r_w \, ,
\eqe
where $H$ is defined by $3 H^2 \defeq \Lambda$ in Eq.\eref{eq:ass.SdS.f.H}, and $\lambda$ is an arbitrary scaling rate. 
The extensivity and intensivity of each state variable of our thermal equilibrium systems should be defined under these fundamental length size scalings as explained in the assumption~2.
\footnote{
As explained in Appendix~B in previous paper~\cite{ref:euclidean.ds}, when we regard $A \defeq 4\,\pi\,r_w^2$ as a state variable of system size, the scaling of system size should be restricted to the homothetic one, which is the spherical scaling due to the spherical symmetry of SdS spacetime. 
The fundamental length size scaling~\eref{eq:ass.scaling} is consistent with this restriction. 
See Appendix~B in previous paper~\cite{ref:euclidean.ds} for details of such restriction.
}

\subsubsection{Euclidean action method and how to obtain state variables}

We need to specify how to get the state variables. 
As noted in the step~2 in assumption~1, thermodynamics of our two thermal equilibrium systems should be constructed in the canonical ensemble. 
Therefore we use the Euclidean action method which is the technique to calculate the partition function of quantum gravity~\cite{ref:euclidean}. 
Indeed, the Euclidean action method has already made successes to obtain the partition function of canonical ensemble for the thermodynamics of single-horizon spacetimes~\cite{ref:euclidean.sch,ref:euclidean.kn,ref:euclidean.sads,ref:euclidean.ds}. 
The key point~3 in Appendix~\ref{app:key} explains how to use the Euclidean action and to define state variables for Schwarzschild thermodynamics. 
Then, referring to the key point~3, we adopt the following assumption:
\begin{ass}[State variables and Euclidean action method] 
Euclidean actions $I_{Eb}$ and $I_{Ec}$ of our two thermal equilibrium systems yield the partition functions of canonical ensembles by Eq.\eref{eq:app.action.Zcl} of Appendix~\ref{app:action}. 
And the free energies $F_b$ and $F_c$ are defined by Eq.\eref{eq:app.action.F}, where the temperatures are defined by Eq.\eref{eq:app.action.T}. 
Then, once $F_b$ and $F_c$ are determined, all state variables of BEH and CEH are defined from $F_b$ and $F_c$ as for thermodynamics of ordinary laboratory systems. 
For example, BEH entropy $S_b$ is defined by $S_b \defeq -\partial F_b/\partial T_b$, where $T_b$ is the temperature of BEH.
\end{ass}

As seen in Eq.\eref{eq:app.action.IE.curved}, the Euclidean action is produced from the Lorentzian action. 
The Lorentzian Einstein-Hilbert action $I_L$ is
\eqb
 I_L \defeq
  \frac{1}{16 \pi}\int_{\mathcal M} dx^4\,\sqrt{-g}\,\left( R - 2\,\Lambda \right)
 + \frac{1}{8\,\pi}\int_{\partial \mathcal M} dx^3\,\sqrt{h}\,K
 + I_0 \, ,
\label{eq:ass.IL}
\eqe
where ${\mathcal M}$ is the spacetime region under consideration, $R$ is the Ricci scalar, $g$ is the determinant of metric, $h$ and $K$ in the second term are respectively the determinant of first fundamental form (induced metric) and the trace of second fundamental form (extrinsic curvature) of the boundary ${\partial \mathcal M}$, and $I_0$ is the integration constant of $I_L$ and called the subtraction term. 
The second term in Eq.\eref{eq:ass.IL} is required to eliminate the second derivatives of metric from the action~\cite{ref:action}. 
The $I_0$ is independent of the metric in ${\mathcal M}$, and does not contribute to the Einstein equation obtained by $\delta I_L = 0$. 
The Einstein equation for SdS spacetime gives the relation $R = 4\,\Lambda$.

When we use the Euclidean action method, it is necessary to specify the integration constant $I_0$. 
It is natural to require that our thermal equilibrium systems for BEH and CEH should reproduce, respectively, the Schwarzschild thermodynamics in the limit $\Lambda \to 0$ and the de~Sitter thermodynamics in the limit $M \to 0$. 
Then the following working hypothesis is naturally required:
\begin{hypo}[Integration constants in Euclidean action]
For the thermal equilibrium system for BEH, the integration constant in Euclidean action is determined with referring to Schwarzschild canonical ensemble formulated by York~\cite{ref:euclidean.sch}. 
For the thermal equilibrium system for CEH, the integration constant in Euclidean action is determined with referring to de~Sitter canonical ensemble formulated in previous paper~\cite{ref:euclidean.ds}.
\end{hypo}

We introduce this working hypothesis as if this is a statement separated from the assumption~3. 
However the determination of integration constant accompanies necessarily the Euclidean action method. 
The working hypothesis~2 is regarded as a part of the assumption~3.

\subsubsection{Effects of external gravitational fields}

By the assumption~3 together with the working hypothesis~2, the concrete functional form of free energies $F_b$ and $F_c$ can be determined as functions of three independent working variables $(M,\Lambda,r_w)$. 
However, since the form of the state variables $X_b$ and $X_c$ have not been specified yet, we can not rearrange $F_b$ and $F_c$ to functions of independent \emph{state variables}, $(T_b,A,X_b)$ and $(T_c,A,X_c)$.

As explained at Eqs.\eref{eq:ass.Fb} and~\eref{eq:ass.Fc}, the CEH (BEH) is regarded as the source of external gravitational field which affects the thermodynamic state of BEH (CEH). 
This means that the state variables $X_b$ and $X_c$ represent, respectively, the thermodynamic effect of CEH's gravity on BEH and that of BEH's gravity on CEH. 
Then it is reasonable to expect that $X_b$ depends on the quantity characterizing the CEH's gravity, and $X_c$ depends on the quantity characterizing the BEH's gravity. 
Moreover, due to the step~3 in assumption~1, $X_b$ and $X_c$ should be measurable for the observer at $r_w$. 
Then, we can offer two candidates for the pair of \emph{dimensionless} characteristic quantities of BEH's and CEH's gravities; 
\begin{itemize}
\item
First candidate pair consists of $\kappa_b r_w$ and $\kappa_c r_w$, where $\kappa_b r_w$ is for BEH's gravity and $\kappa_c r_w$ is for CEH's gravity. This pair means that both of BEH's and CEH's gravities are characterized by three quantities $(M , \Lambda , r_w)$, since $\kappa_b$ and $\kappa_c$ depend on $M$ and $\Lambda$. 
\item
Second candidate pair consists of $M/r_w$ and $H r_w$, where $M/r_w$ is for BEH's gravity and $H r_w$ is for CEH's gravity. This pair means that the BEH's gravity is characterized by $(M , r_w)$, and the CEH's gravity is characterized by $(\Lambda , r_w)$. 
\end{itemize}
Here, purely logically, we can consider the ``inverse'' pair of second one, where $H r_w$ is for BEH's gravity and $M/r_w$ is for CEH's gravity. 
This means that the BEH's gravity is characterized by $(\Lambda , r_w)$, and the CEH's gravity is characterized by $(M , r_w)$. 
However this is physically unacceptable, since we do not expect that the BEH does not depend on $M$ and the CEH does not depend on $\Lambda$. 
Therefore, the reasonable candidates for the pair of characteristic quantities of BEH's and CEH's gravities are the two candidates listed above. 
Then, $X_b$ should be a function of $( \kappa_c  \,,\, r_w )$ or $( H \,,\, r_w )$, and $X_c$ should be a function of $( \kappa_b \,,\, r_w )$ or $( M \,,\, r_w )$.

On the other hand, as will be mathematically verified in Secs.\ref{sec:beh.3} and~\ref{sec:ceh.3}, the state variables $X_b$ and $X_c$ are the extensive variables and proportional to $r_w^2$. 
The proportionality to $r_w^2$ is consistent with the scaling law of extensive variables denoted in assumption~2. 
And, according to the previous paragraph, the factor of proportionality should be a function of the characteristic quantity of BEH's or CEH's gravity. 
Although the verification of the extensivity of $X_b$ and $X_c$ are shown later, we accept it in the following assumption~4 for the simplicity of our discussion.

From the above, it is reasonable to adopt the following assumption:
\begin{ass}[Extensive variable of ``external field''] 
The state variables $X_b$ and $X_c$ in Eqs.\eref{eq:ass.Fb} and~\eref{eq:ass.Fc} are the extensive variables. 
(This will be verified in Secs.{\ref{sec:beh.3}} and {\ref{sec:ceh.3}}). 
Then, there are two candidates for the functional forms of $X_b$ and $X_c$. 
One of them is based on the quantities $(\kappa_b r_w \,,\, \kappa_c r_w)$:
\eqb
\label{eq:ass.X.1}
 X_b = r_w^2\, \Psi_b(\kappa_c r_w) \quad,\quad X_c = r_w^2\,\Psi_c(\kappa_b r_w) \, ,
\eqe
where $\Psi_b$ and $\Psi_c$ are arbitrary functions of single argument, whose functional forms are not specified at present. 
Another candidate of $X_b$ and $X_c$ is based on the quantities $(M/r_w \,,\, H\,r_w)$:
\eqb
\label{eq:ass.X.2}
 X_b = r_w^2\, \Psi_b(H\,r_w) \quad,\quad X_c = r_w^2\,\Psi_c(M/r_w) \, .
\eqe
At least for the present author, no criterion to choose one of these candidates is found, and the way for determining the functional forms of $\Psi_b$ and $\Psi_c$ are also unknown. 
An obvious constraint on $\Psi_b$ and $\Psi_c$ is that they never be constant to make $X_b$ and $X_c$ independent of the state variable of system size $A \defeq 4 \pi r_w^2$.
\end{ass}

In Sec.\ref{sec:sd.1}, we will make some comments on the issue which of Eqs.\eref{eq:ass.X.1} and~\eref{eq:ass.X.2} is valid. 
Those comments will suggest that Eq.\eref{eq:ass.X.1} may be appropriate, but we do not have mathematical verification to choose Eq.\eref{eq:ass.X.1} as the general form of $X_b$ and $X_c$. 
Therefore, to retain the logical strictness of this paper, we list the two possibilities~\eref{eq:ass.X.1} and~\eref{eq:ass.X.2} in the assumption~4.

From the above, we recognize that the minimal set of assumptions for ``consistent'' thermodynamics of our two thermal equilibrium systems should be composed of four assumptions. 
However, the determination of functions $\Psi_b$ and $\Psi_c$ remains as a future task and we can not find concrete functional forms of them. 
Although the state variables $X_b$ and $X_c$ are not specified in this paper, the existence of them enables us to examine the validity of entropy-area law in SdS spacetime as shown in Secs.~\ref{sec:beh} and~\ref{sec:ceh}.

\section{Euclidean actions}
\label{sec:action}

Referring to the assumption~3 and working hypothesis~2, we calculate Euclidean actions for the two thermal equilibrium systems for BEH and CEH constructed in the assumption~1.

\subsection{Euclidean action for BEH}

Euclidean space of thermal equilibrium system for BEH is obtained by the Wick rotations $t \to -i\,\tau$ in the static chart and $\eta_b \to -i\,\omega_b$ in the semi-global black hole chart. 
These Wick rotations are equivalent, because the coordinate transformation~\eref{eq:ass.trans.beh}, $\eta_b = e^{\kappa_b\, r^{\ast}}\sinh\left(\kappa_b\, t\right)$, implies that the imaginary time $\omega_b$ in the semi-global chart is defined by $\omega_b \defeq e^{\kappa_b\, r^{\ast}}\sin\left(\kappa_b\, \tau\right)$, where $\tau$ is the imaginary time in the static chart. 
Euclidean metric in the static chart is
\eqb
 ds_E^2 = f(r)\,d\tau^2 + \frac{dr^2}{f(r)} + r^2\,d\Omega^2 \, .
\label{eq:action.SdSE.static} \\
\eqe
Euclidean metric in the semi-global black hole chart is
\eqb
 ds_E^2 = \Upsilon_b(r)\,\left[\,d\omega_b^2 + d\chi_b^2 \right] + r^2\,d\Omega^2 \, ,
\label{eq:action.SdSE.global.beh}
\eqe
where $\Upsilon_b$ is defined in Eq.\eref{eq:ass.upsilon.beh}. 
About the semi-global chart, we get from the coordinate transformation~\eref{eq:ass.trans.beh},
\eqb
 \omega_b^2 + \chi_b^2 =
  \left( \frac{r}{r_b} - 1 \right)\,\left( 1 - \frac{r}{r_c} \right)^{-\kappa_b/\kappa_c}
  \left( \frac{r}{r_b + r_c} + 1 \right)^{-1+\kappa_b/\kappa_c} \, .
\label{eq:action.SdSE.radial}
\eqe
Then, because the thermal equilibrium system for BEH is the region $D_b$, $r_b < r < r_w$, in Lorentzian SdS spacetime, we find the topology of Euclidean space of thermal equilibrium system for BEH is $D^2\times S^2$. 
Because $\omega_b^2 + \chi_b^2 = 0$ at $r = r_b$, the center of $D^2$-part is at the BEH $r = r_b$, and the boundary of $D^2$-part is at the heat wall $r = r_w$. 
The topology of heat wall boundary is $S^1 \times S^2$, where $S^1$ is along the $\tau$-direction.

Because the Lorentzian SdS spacetime is regular at $r_b$, the Euclidean space is also regular at $r_b$.
To examine the regularity of Euclidean space at $r_b$, we make use of the static chart~\eref{eq:action.SdSE.static}. 
Let us define a coordinate $x_b$ and a function $\gamma(x_b)$ by
\eqb
 x_b^2 \defeq r - r_b \quad,\quad
 \gamma(x_b) \defeq \sqrt{f(r_b + x_b^2)} \, .
\eqe
We get $\gamma'(x_b) \defeq d\gamma(x_b)/dx_b = [ x_b/\gamma(x_b) ]\,df(r)/dr$, from which we find
\eqb
 \lim_{x_b \to 0}\gamma'(x_b)
  = \left.\frac{d C}{dr}\right|_{r_b}\,\lim_{x_b\to 0}\frac{x_b}{\gamma(x_b)}
  = 2\,\kappa_b\,\frac{1}{\displaystyle \lim_{x_b \to 0}\gamma'(x_b)} \, .
\eqe
This means $\gamma'(0) = \sqrt{2\,\kappa_b}$, and near the BEH, $f \simeq \left[ \gamma(0) + \gamma'(0)\,x \right]^2 = 2\,\kappa\,x_b^2$. 
Therefore the Euclidean metric near BEH is
\eqb
 ds_E^2 \simeq
 \frac{2}{\kappa_b}\,\left[\, x_b^2\,d(\kappa_b\, \tau)^2 + dx_b^2 \,\right] + r^2\,d\Omega^2 \, .
\eqe
It is obvious that the Euclidean space is regular at BEH if the imaginary time has the period $\beta_b$ defined by
\eqb
 0 \le \tau < \beta_b \defeq \frac{2\,\pi}{\kappa_b} \, .
\label{eq:action.period.beh}
\eqe
Throughout our discussion, $\tau$ has the period $\beta_b$ in the Euclidean space of thermal equilibrium system for BEH.

Now let us proceed to the calculation of the Euclidean action $I_{Eb}$ of thermal equilibrium system for BEH. 
Following the working hypothesis~2, we use the same integration constant $I_0$ as Schwarzschild canonical ensemble~\cite{ref:euclidean,ref:euclidean.sch}, which gives us
\eqb
\label{eq:action.I0.beh}
 I_0 \defeq - I_L^{\rm (flat)}
     = - \frac{1}{8 \pi}\,\int_{\partial \mathcal M}\,dx^3 \,
         \sqrt{h}\,K^{\rm (flat)} \, ,
\eqe
where $I_L^{\rm (flat)}$ is the Lorentzian Einstein-Hilbert action for flat spacetime which reduces to only the surface term (the second term in Eq.\eref{eq:ass.IL} ) due to $R = 0$ and $\Lambda = 0$ for flat spacetime, and $K^{\rm (flat)}$ is the trace of second fundamental form of $\partial \mathcal M$ in flat spacetime~\cite{ref:euclidean,ref:euclidean.sch}. 
Here note that, the integral element $\sqrt{h}$ in $I_L^{\rm (flat)}$ should be given by that of SdS spacetime when $I_L^{\rm (flat)}$ is used as the integration constant of action integral of SdS spacetime, because the background spacetime on which the integral in $I_L^{\rm (flat)}$ is calculated is SdS spacetime.

For our thermal equilibrium system for BEH in SdS spacetime, the region ${\mathcal M}$ in $I_L$ is $D_b$, $r_b < r < r_w$, and its boundary $\partial D_b$ is at $r_w$. 
There is another boundary at $r_b$ in the Lorentzian region $D_b$. 
However we do not need to consider it, because the points at $r_b$ in the Euclidean space do not form a boundary but are the regular points when $\tau$ has the period~\eref{eq:action.period.beh}. 
Then the first fundamental form $h_{i j}$ ($i, j = 0, 2, 3$) of $\partial D_b$ in the static chart is
\eqb
 \left. ds^2 \right|_{r = r_w} = h_{i j}\, dx^i \, dx^j
 = - f_w\,dt^2 + r_w^2\,d\Omega^2 \, ,
\eqe
where
\eqb
 f_w \defeq f(r_w) = 1 - \frac{2 M}{r_w} - H^2\,r_w^2 \, .
\label{eq:action.fw}
\eqe
Here, since $D_b$ is the region enclosed by BEH and heat wall, the direction of unit normal vector $n^{\mu}$ to $\partial D_b$ is pointing towards CEH, $n^{\mu} \propto \partial_r$. 
Then the second fundamental form of $\partial D_b$ in the static chart is
\eqb
\label{eq:action.Kij.beh}
 K_{i j} =
 \sqrt{f_w}\, 
 {\rm diag.}\left[\, - \frac{M}{r_w^2} + H^2\,r_w \,,\, r_w \,,\, r_w\,\sin^2\theta \, \right] \, ,
\eqe
where diag. means the diagonal matrix form.

On the other hand, the second fundamental form $K_{ij}^{\rm (flat)}$ of a spherically symmetric timelike hypersurface of radius $r_w$ in flat spacetime is given by setting $M = 0$ and $H = 0$ in Eq.\eref{eq:action.Kij.beh},
\eqb
 K_{i j}^{\rm (flat)} =
 {\rm diag.}\left[\, 0 \,,\, r_w \,,\, r_w\,\sin^2\theta \, \right] \, .
\eqe
This gives $K^{\rm (flat)} \defeq h^{ij}\,K_{ij}^{\rm (flat)} = 2\,r_w^{-1}$.

From the above, applying the Wick rotation $t \to -i \tau$ to the Lorentzian action $I_L$ in Eq.\eref{eq:ass.IL}, we obtain the Euclidean action $I_{E b}$ of the thermal equilibrium system for BEH via Eq.\eref{eq:app.action.IE.curved},
\eqb
\label{eq-setting.IEb}
\begin{split}
 I_{Eb}
 &= \frac{3\,H^2}{8\,\pi} \int_{D_{Eb}}dx_E^4\,\sqrt{g_E}
    + \frac{1}{8\,\pi}\int_{\partial D_{Eb}} dx_E^3 \,
                      \sqrt{h_E}\,\left(\,K_E - K_E^{\rm (flat)}\,\right) \\
 &= \frac{\beta_b}{2}\,
    \left[\, 3\,M - r_b + 2\,r_w \left( f_w - \sqrt{f_w} \right) \,\right] \, ,
\end{split}
\eqe
where the relation for SdS spacetime $R = 4\,\Lambda = 12\,H^2$ is used in the first equality, the relation $M - H^2 r_b^3 = 3 M - r_b$ due to $f(r_b) = 0$ is used in the second equality, $Q_E$ is the quantity $Q$ evaluated on Euclidean space, and $D_{Eb}$ is the Euclidean region denoted by $0 \le \tau < \beta_b$ , $r_b \le r \le r_w$ , $0 \le \theta \le \pi$ and $0 \le \varphi < 2\,\pi$. 
This $I_{Eb}$ corresponds to $I_E[g_{E\,cl}]$ in Eq.\eref{eq:app.action.F} of Appendix~\ref{app:action}, which yields the partition function of our thermal equilibrium system for BEH.

Note that $I_{Eb}$ should reproduce the Euclidean action of Schwarzschild canonical ensemble as required in the working hypothesis~2. 
To check if this is satisfied, take the limit $\Lambda \to 0$, 
\eqb
 \lim_{\Lambda \to 0} I_{Eb} = 4 \pi M\,\left[\, M - 2\,r_w \left( f_w - \sqrt{f_w} \right) \,\right]_{\Lambda=0} \, .
\eqe
This coincides with the Euclidean action of Schwarzschild canonical ensemble formulated by York~\cite{ref:euclidean.sch}.

\subsection{Euclidean action for CEH}

Euclidean space of thermal equilibrium system for CEH is obtained by the Wick rotations $t \to -i\,\tau$ in the static chart and $\eta_c \to -i\,\omega_c$ in the semi-global cosmological chart. 
These Wick rotations are equivalent, because the coordinate transformation~\eref{eq:ass.trans.ceh}, $\eta_c = e^{-\kappa_c\, r^{\ast}}\sinh\left(\kappa_c\, t\right)$, implies that the imaginary time $\omega_c$ in the semi-global chart is defined by $\omega_c \defeq e^{-\kappa_c\, r^{\ast}}\sin\left(\kappa_c\, \tau\right)$, where $\tau$ is the imaginary time in the static chart. 
Euclidean metric in the static chart is given by Eq.\eref{eq:action.SdSE.static}. 
Euclidean metric in the semi-global cosmological chart is $ds_E^2 = \Upsilon_c(r)\,\left[\,d\omega_c^2 + d\chi_c^2 \right] + r^2\,d\Omega^2$, where $\Upsilon_c$ is defined in Eq.\eref{eq:ass.upsilon.ceh}. 
About the semi-global chart, we get from the coordinate transformation~\eref{eq:ass.trans.ceh},
\eqb
 \omega_c^2 + \chi_c^2 =
  \left( \frac{r}{r_b} - 1 \right)^{-\kappa_c/\kappa_b}\,\left( 1 - \frac{r}{r_c} \right)
  \left( \frac{r}{r_b + r_c} + 1 \right)^{-1+\kappa_c/\kappa_b} \, .
\eqe
Then, because the thermal equilibrium system for CEH is the region $D_c$, $r_w < r < r_c$, in Lorentzian SdS spacetime, we find the topology of Euclidean space of thermal equilibrium system for CEH is $D^2\times S^2$. 
Because $\omega_c^2 + \chi_c^2 = 0$ at $r = r_c$, the center of $D^2$-part is at the CEH $r = r_c$, and the boundary of $D^2$-part is at the heat wall $r = r_w$. 
The topology of heat wall boundary is $S^1 \times S^2$, where $S^1$ is along the $\tau$-direction.

Because the Lorentzian SdS spacetime is regular at $r_c$, the Euclidean space is also regular at $r_c$.
Using the static chart~\eref{eq:action.SdSE.static} and defining a coordinate $x_c$ and a function $\gamma(x_c)$ by $x_c^2 \defeq r_c - r$ and $\gamma(x_c) \defeq \sqrt{f(r_c - x_c^2)}$, we obtain the Euclidean metric near CEH,
\eqb
 ds_E^2 \simeq
 \frac{2}{\kappa_c}\,\left[\, x_c^2\,d(\kappa_c\, \tau)^2 + dx_c^2 \,\right] + r^2\,d\Omega^2 \, .
\eqe
It is obvious that the Euclidean space is regular at CEH if the imaginary time has the period $\beta_c$ defined by
\eqb
 0 \le \tau < \beta_c \defeq \frac{2\,\pi}{\kappa_c} \, .
\label{eq:action.period.ceh}
\eqe
Throughout our discussion, $\tau$ has the period $\beta_c$ in the Euclidean space of thermal equilibrium system for CEH.

Now we proceed to the calculation of Euclidean action $I_{Ec}$ of the thermal equilibrium system for CEH. 
Lorentzian action is defined in Eq.\eref{eq:ass.IL} for the spacetime region $D_c$, $r_w < r < r_c$.
The calculation of Euclidean action $I_{Ec}$ is parallel to that of $I_{Eb}$ except for the direction of unit normal vector $n^{\mu}$ to $\partial D_c$ and the integration constant $I_0$ in $I_L$. 
Concerning the vector $n^{\mu}$, since $D_c$ is the region enclosed by CEH and heat wall, the direction of $n^{\mu}$ is pointing towards BEH, $n^{\mu} \propto - \partial_r$.

Concerning the integration constant, following the working hypothesis~2, we determine the integration constant $I_0$ for CEH with referring to the de~Sitter canonical ensemble formulated in previous paper~\cite{ref:euclidean.ds}. 
At the limit $M \to 0$, the term $I_0$ should reduce to the integration constant $I_0^{\rm (dS)}$ of the de~Sitter canonical ensemble, 
\eqb
 I_0(M=0) = I_0^{\rm (dS)}
 \defeq \left(\frac{1}{H\,r_w} - 1 \right)\,\sqrt{f_w(M=0)} \, I_L^{\rm (flat)} \, ,
\eqe
where $I_L^{\rm (flat)}$ is the action of flat spacetime used in Eq.\eref{eq:action.I0.beh}. 
Note that the CEH radius in de~Sitter spacetime is $H^{-1}$, and the factor $H\,r_w$ is the ratio of heat wall radius $r_w$ to CEH radius. 
Hence we set $I_0$ for the CEH in SdS spacetime,
\eqb
\label{eq:action.I0.ceh}
 I_0 \defeq \left(\frac{r_c}{r_w} - 1 \right)\,\sqrt{f_w}\,I_L^{\rm (flat)} \, ,
\eqe
where it should be emphasized that the signature of $K^{\rm (flat)}$ in the integrand of $I_L^{\rm (flat)}$ (shown in Eq.\eref{eq:action.I0.beh}) should be reversed, because the direction of normal vector $n^{\mu}$ is reversed as mentioned in previous paragraph.

Then we obtain the Euclidean action $I_{Ec}$ of our thermal equilibrium system for CEH via Eqs.\eref{eq:ass.IL} and~\eref{eq:app.action.IE.curved},
\eqb
\label{eq:action.IEc}
 I_{Ec}
 = - \frac{\beta_c}{2}\,\left[\, 3\,M - r_c + 2\,r_c\,f_w \,\right] \, ,
\eqe
where the relation $M - H^2 r_c^3 = 3 M - r_c$ due to $f(r_c) = 0$ is used, and the overall minus signature comes from the direction of normal vector $n^{\mu}$ to $\partial D_c$. 
This $I_{Ec}$ corresponds to $I_E[g_{E\,cl}]$ in Eq.\eref{eq:app.action.F}, which yields the partition function of our thermal equilibrium system for CEH.

Note that $I_{Ec}$ should reproduce the Euclidean action of de~Sitter canonical ensemble as required in the working hypothesis~2. 
To check if this is satisfied, take the limit $M \to 0$, 
\eqb
 \lim_{M \to 0} I_{Ec} = - \frac{\pi}{H^2}\,\left[\, 1 - 2\,(H r_w)^2 \,\right] \, .
\eqe
This coincides with the Euclidean action of de~Sitter canonical ensemble formulated in previous paper~\cite{ref:euclidean.ds}.

\section{Black hole event horizon}
\label{sec:beh}

We examine whether the entropy-area law holds for ``consistent'' thermodynamics of our thermal equilibrium system for BEH.

\subsection{Temperature and free energy of BEH}
\label{sec:beh.1}

By the assumption~3, the temperature $T_b$ of BEH is defined by Eq.\eref{eq:app.action.T} of Appendix~\ref{app:action}, which relates $T_b$ to the proper length in the imaginary time direction at the boundary $\partial D_b$ (the direction along $S^1$ part of boundary topology $S^1 \times S^2$ in Euclidean space),
\eqb
 T_b \defeq \left[\, \int_0^{\beta_b}\,\sqrt{f_w}\,d\tau \,\right]^{-1}
     = \frac{\kappa_b}{2\,\pi\,\sqrt{f_w}} \, ,
\label{eq:beh.Tb}
\eqe
where $\beta_b$ is the imaginary time period~\eref{eq:action.period.beh} and $f_w$ is in Eq.\eref{eq:action.fw}. 
Under the length size scaling~\eref{eq:ass.scaling}, this temperature is scaled as $T_b \to \lambda^{-1}\,T_b$. 
Therefore, by the assumption~2, $T_b$ is an intensive state variable of BEH.

Note that this $T_b$ coincides with the Hawking temperature of BEH derived originally by Gibbons and Hawking~\cite{ref:temperature}, and the factor $\sqrt{f_w}$ is the so-called Tolman factor~\cite{ref:tolman} which expresses the gravitational redshift affecting the Hawking radiation propagating from BEH to observer at $r_w$. 
Therefore this $T_b$ is the temperature measured by the observer at heat wall.

By the assumption~2, the extensive state variable of system size for our thermal equilibrium system is the surface area of heat wall,
\eqb
 A \defeq 4\,\pi\,r_w^2 \, .
\eqe

By the assumption~3, the free energy $F_b$ of BEH in Eq.\eref{eq:ass.Fb} is defined by Eq.\eref{eq:app.action.F},
\eqb
 F_b(T_b,A,X_b) \defeq - T_b\,I_{Eb} = r_w - r_w\,\sqrt{f_w} - \frac{3\,M - r_b}{2\,\sqrt{f_w}} \, .
\label{eq:beh.Fb}
\eqe
Under the length size scaling~\eref{eq:ass.scaling}, this free energy satisfies the scaling law of thermodynamic functions, $F_b \to \lambda\,F_b$. 
As discussed at Eq.\eref{eq:ass.Fb}, $F_b$ is regarded as a function of three independent state variables $T_b$, $A$ and the state variable $X_b$ of CEH's gravitational effect on BEH. 
However, since $X_b$ is not specified as mentioned in the assumption~4, the form of $F_b$ as a function of $( T_b\,,\,A\,,\,X_b )$ remains unknown. 
Instead, Eq.\eref{eq:beh.Fb} shows $F_b$ as a function of independent parameters $( M\,,\,H\,,\,r_w )$.

Let us verify that the free energy $F_b$ is a function of three independent state variables. 
We use the reductive absurdity: 
Assume that only two (not three) state variables are independent. 
This assumption means that, as for the ordinary non-magnetized gases, $F_b$ is a function of $T_b$ and $A$, $F_b(T_b,A)$. 
Here it is obvious from Eq.\eref{eq:beh.Tb} that $T_b$ depends on three parameters $(M,H,r_w)$, while $A$ depends only on $r_w$. 
Then, via Eq.\eref{eq:app.formulas.3rd} of Appendix~\ref{app:formulas}, a mathematical relation $(\partial_M F_b)/(\partial_M T_b) = (\partial_H F_b)/(\partial_H T_b)$ must hold if $F_b$ is a function of $(T_b , A)$. 
However we find this relation does not hold, $(\partial_M F_b)/(\partial_M T_b) \neq (\partial_H F_b)/(\partial_H T_b)$, via Eq.\eref{eq:beh.dFb_dTb} shown below. 
Hence the assumption of two independent state variables is denied by the reductive absurdity. 
Now the working hypothesis~1, which assumes $F_b$ to be a function of three independent state variables, is verified.

To support the discussion in previous paragraph and for later use, we show some differentials:
\seqb
\eqab
 \pd{T_b}{M} &=&
  \frac{1}{2 \pi r_b^2 f_w^{3/2}}
  \left[ \frac{r_b}{2 r_w} \left( 1 - 3 H^2 r_b^2 \right)
       - \frac{1 + 3 H^2 r_b^2}{1 - 3 H^2 r_b^2}\, f_w \right]
\label{eq:beh.dTb/dM} \, , \\
 \pd{T_b}{H} &=&
  \frac{1}{4 \pi H r_b f_w^{3/2}}
  \left[ \left( 1 - \frac{2 M}{r_w} \right) \left( 1 - 3 H^2 r_b^2 \right)
       - \frac{1 + 3 H^2 r_b^2}{1 - 3 H^2 r_b^2}\,\, \frac{2 M f_w}{r_b} \right]
\label{eq:beh.dTb/dH} ,
\eqae
and
\eqab
\label{eq:beh.dFb/dM}
 \pd{F_b}{M} &=&
  \frac{1}{2 f_w^{3/2}}
  \left[ - \frac{3 M - r_b}{r_w} + \frac{1 + 3 H^2 r_b^2}{1 - 3 H^2 r_b^2}\, f_w \right] \, , \\
\label{eq:beh.dFb/dH}
 \pd{F_b}{H} &=&
  \frac{1}{2 f_w^{3/2}}
  \left[ \left( 2 r_w + r_b - 7 M - 2 H^2 r_w^3 \right) H r_w^2
       + \frac{2 H r_b^3 f_w}{1 - 3 H^2 r_b^2} \right] \, ,
\eqae
\seqe
where the differentials in Eqs.\eref{eq:ass.diff.rb} and~\eref{eq:ass.diff.kappab} are used. 
Then we get
\seqb
\label{eq:beh.dFb_dTb}
\eqab
 \pd{F_b}{M} &=& - \pi\,r_b^2\,\pd{T_b}{M}
\label{eq:beh.dFb/dM.dTb/dM} \, , \\
 \pd{F_b}{H} &\not\propto& \pi\,r_b^2\,\pd{T_b}{H}
\label{eq:beh.dFb/dH.dTb/dH} \, ,
\eqae
\seqe
where the definition of $r_b$, $f(r_b) = 0$, is used in the first relation. 
These relations are used in examining the entropy-area law in next subsection.

\subsection{Entropy of BEH}
\label{sec:beh.2}

By the assumption~3, the entropy $S_b$ of BEH is defined as the thermodynamic conjugate variable to $T_b$,
\eqb
 S_b \defeq - \pd{F_b(T_b,A,X_b)}{T_b} \, .
\label{eq:beh.Sb.def}
\eqe
Under the length size scaling~\eref{eq:ass.scaling}, this entropy satisfies the extensive scaling law, $S_b \to \lambda^2\,S_b$. 
By Eq.\eref{eq:app.formulas.2parameters} of Appendix~\ref{app:formulas}, $S_b$ is rearranged to
\eqb
 S_b =
 -\frac{(\partial_M F_b)\,(\partial_H X_b) - (\partial_H F_b)\,(\partial_M X_b)}
       {(\partial_M T_b)\,(\partial_H X_b) - (\partial_H T_b)\,(\partial_M X_b)} \, .
\label{eq:beh.Sb}
\eqe
Then we get by Eq.\eref{eq:beh.dFb/dM.dTb/dM},
\eqb
 S_b =
 \frac{\pi\,r_b^2 \,(\partial_M T_b)\,(\partial_H X_b) + (\partial_H F_b)\,(\partial_M X_b)}
      {(\partial_M T_b)\,(\partial_H X_b) - (\partial_H T_b)\,(\partial_M X_b)} \, .
\label{eq:beh.Sb.2}
\eqe
This denotes the following: If $\partial_M X_b \equiv 0$, then the entropy-area law $S_b \equiv \pi\,r_b^2$ holds. 
However, if $\partial_M X_b \not\equiv 0$, then Eq.\eref{eq:beh.dFb/dH.dTb/dH} together with Eq.\eref{eq:beh.Sb.2} imply that the entropy-area law breaks down $S_b \not\equiv \pi\,r_b^2$. 
Therefore we find that the entropy-area law holds if and only if $\partial_M X_b \equiv 0$.

In summary, although the validity of entropy-area law for BEH can not be judged at present, we can clarify the issue on the entropy-area law for BEH: 
\emph{The necessary and sufficient condition to ensure the entropy-area law for BEH is that the BEH is in thermal equilibrium and the state variable $X_b$ satisfies $\partial_M X_b \equiv 0$. 
If the CEH's gravitational effect on BEH, $X_b$, is characterized by the CEH's quantity $\kappa_c r_w$ as shown in Eq.\eref{eq:ass.X.1}, then $\partial_M X_b \not\equiv 0$ and the entropy-area law breaks down for BEH. 
If the CEH's gravitational effect $X_b$ is characterized by $H r_w$ as shown in Eq.\eref{eq:ass.X.2}, then $\partial_M X_b \equiv 0$ and the entropy-area law holds.
The validity of entropy-area law for BEH will be judged by revealing which of $\kappa_c$ or $H$ is appropriate as the characteristic quantity of CEH's gravity.}

\subsection{Thermodynamic consistency of BEH}
\label{sec:beh.3}

The remaining part of Sec.\ref{sec:beh} is for the ``thermodynamic consistency'' of our thermodynamics of BEH under the minimal set of assumptions. 
Since the concrete form of $X_b$ is not specified, the following discussions and calculations are very formal. 
However we can imply that the thermodynamic consistency is satisfied. 
Moreover, it is also verified that $X_b$ is an extensive variable and proportional to $r_w^2$.

Following the assumption~3, the internal energy $E_b$ of BEH can be defined by the argument of statistical mechanics, 
\eqb
\label{eq:beh.Fb.Eb}
 E_b \defeq - \left.\pd{\ln Z_{cl}}{(1/T_b)}\right|_{A, X_b = \mbox{const.}}
 = \pd{(F_b/T_b)}{(1/T_b)} = F_b + T_b\,S_b \, ,
\eqe
where Eq.\eref{eq:app.action.F} of Appendix~\ref{app:action} is used in the second equality, and the definition of $S_b$ in Eq.\eref{eq:beh.Sb.def} is used in the third equality. 
Under the length size scaling~\eref{eq:ass.scaling}, this $E_b$ satisfies the scaling law of thermodynamic functions, $E_b \to \lambda\,E_b$. 
The third equality in Eq.\eref{eq:beh.Fb.Eb}, $E_b = F_b + T_b\,S_b$, is regarded as the Legendre transformation between $F_b$ and $E_b$, which determines $E_b$ to be a function of $(S_b,A,X_b)$. 
On the other hand, in thermodynamic argument, the internal energy is the thermodynamic function which is regarded as a function of only extensive state variables. 
Hence, it is verified that $X_b$ must be an extensive state variable. 
(The proportionality of $X_b$ to $r_w^2$ will also be shown mathematically at the end of this subsection.)

Next, in order to see the first law, we need the intensive state variables which are thermodynamically conjugate to $A$ and $X_b$. 
By the assumption~3, the conjugate state variable is defined by the appropriate differential of a thermodynamic function. 
In an analogy with the ordinary pressure of ordinary gases, the \emph{surface pressure} (at the heat wall) $\sigma_b$ is defined formally as
\seqb
\eqab
\label{eq:beh.sigmab.def}
 \sigma_b && \defeq - \pd{F_b(T_b,A,X_b)}{A} \\
&&
\label{eq:beh.sigmab}
\begin{aligned}
 &= -\frac{1}{8 \pi r_w}\,
      \frac{1}{(\partial_M T_b)\,(\partial_H X_b) - (\partial_M X_b)\,(\partial_H T_b)} \\
 & \quad \times
    \Bigl[\, \left\{\, (\partial_H F_b)\,(\partial_{r_w} T_b)
                     - (\partial_{r_w} F_b)\,(\partial_H T_b)\, \right\}\,(\partial_M X_b)\\
 & \qquad\quad + \left\{\, (\partial_{r_w} F_b)\,(\partial_M T_b)
                     - (\partial_M F_b)\,(\partial_{r_w} T_b)\, \right\}\,(\partial_H X_b)\\
 & \qquad\quad + \left\{\, (\partial_M F_b)\,(\partial_H T_b)
                     - (\partial_H F_b)\,(\partial_M T_b)\, \right\}\,(\partial_{r_w} X_b)
    \,\,\, \Bigr] \, ,
\end{aligned}
\eqae
\seqe
where Eq.\eref{eq:app.formulas.3parameters} of Appendix~\ref{app:formulas} is used in the second equality. 
Under the length size scaling~\eref{eq:ass.scaling}, this $\sigma_b$ satisfies the intensive scaling law, $\sigma_b \to \lambda^{-1} \sigma_b$. 
The state variable given by the same definition with $\sigma_b$ appears also in single-horizon thermodynamics~\cite{ref:euclidean.sch,ref:euclidean.kn,ref:euclidean.ds} to ensure the thermodynamic consistency. 
(See Appendix~B in previous paper~\cite{ref:euclidean.ds} for thermodynamic meanings of $A$ and $\sigma_b$.)

The intensive state variable $Y_b$ conjugate to $X_b$ is defined formally as
\eqb
\label{eq:beh.Yb}
 Y_b \defeq \pd{ F_b(T_b , A , X_b)}{X_b}
 =
 \frac{(\partial_M F_b)\,(\partial_H T_b) - (\partial_H F_b)\,(\partial_M T_b)}
      {(\partial_M X_b)\,(\partial_H T_b) - (\partial_H X_b)\,(\partial_M T_b)} \, ,
\eqe
where Eq.\eref{eq:app.formulas.2parameters} of Appendix~\ref{app:formulas} is used in the second equality. 
Under the length size scaling~\eref{eq:ass.scaling}, when $X_b$ is scaled as an extensive variable $X_b \to \lambda^2\,X_b$, then this $Y_b$ satisfies the intensive scaling law, $Y_b \to \lambda^{-1}\,Y_b$.

Then by definitions of $S_b$, $\sigma_b$ and $Y_b$, we get
\eqb
 dF_b(T_b,A,X_b) = - S_b \, dT_b - \sigma_b \, dA + Y_b \, dX_b \, .
\eqe
The first law follows this relation via the Legendre transformation in Eq.\eref{eq:beh.Fb.Eb},
\eqb
 dE_b(S_b,A,X_b) = T_b \, dS_b - \sigma_b \, dA + Y_b \, dX_b \, .
\eqe

Concerning the internal energy, the Euler relation is interesting from the point of view of thermodynamics, because it gives a restriction on the form of state variables. 
By the scaling laws of extensive variable and thermodynamic function, we get
\seqb
\eqb
 \lambda\, E_b(S_b \,,\, A \,,\, X_b) = E_b(\lambda^2 S_b \,,\, \lambda^2 A \,,\, \lambda^2 X_b) \, .
\label{eq:beh.euler.1}
\eqe
This denotes that $E_b(S_b,A,X_b)$ is the homogeneous expression of degree $1/2$. 
Operating the differential $\partial_{\lambda}$ on Eq.\eref{eq:beh.euler.1}, we get
\eqb
 \frac{1}{2}\, E_b(S_b \,,\, A \,,\, X_b) = T_b\,S_b - \sigma_b\,A + Y_b\,X_b \, .
\label{eq:beh.euler.2}
\eqe
\seqe
This relation~\eref{eq:beh.euler.2} is obtained from the scaling behavior~\eref{eq:beh.euler.1}. 
Furthermore by the well-known \emph{Euler's theorem on the homogeneous expression}, the scaling behavior~\eref{eq:beh.euler.1} is also obtained from the relation~\eref{eq:beh.euler.2} (which is proven by the vanishing differential $\partial_{\lambda}[\,\lambda^{-1} E_b(\lambda^2 S_b,\lambda^2 A,\lambda^2 X_b)\,] = 0$). 
Hence Eqs.\eref{eq:beh.euler.1} and~\eref{eq:beh.euler.2} are equivalent. 
As shown below, we find the Euler relation~\eref{eq:beh.euler.2} is consistent with the assumption~4:

By the Legendre transformation in Eq.\eref{eq:beh.Fb.Eb} and the Euler relation~\eref{eq:beh.euler.2}, we get a relation, $F_b = T_b\,S_b - 2\,\sigma_b\,A + 2\,Y_b\,X_b$. 
Then substituting Eqs.\eref{eq:beh.Sb}, \eref{eq:beh.sigmab} and~\eref{eq:beh.Yb} into this relation, we obtain
\eqb
 k_1\,\pd{X_b}{M} + k_2\,\pd{X_b}{H} + r_w\,k_3\,\pd{X_b}{\,r_w}
 = 2\,k_3\,X_b \, ,
\label{eq:beh.pde.k}
\eqe
where
\seqb
\eqab
\label{eq:beh.k1}
 k_1 &\defeq& - F_b\,\left( \partial_H T_b \right) - T_b\,\left( \partial_H F_b \right)\\
 &&\quad - r_w\,\left[\, \left( \partial_H F_b \right)\,\left( \partial_{r_w} T_b \right)
                     - \left( \partial_{r_w} F_b \right)\,\left( \partial_H T_b \right)
                \,\right]
\nonumber \\
\label{eq:beh.k2}
 k_2 &\defeq& F_b\,\left( \partial_M T_b \right) + T_b\,\left( \partial_M F_b \right) \\
 &&\quad - r_w\,\left[\, \left( \partial_{r_w} F_b \right)\,\left( \partial_M T_b \right)
                     - \left( \partial_M F_b \right)\,\left( \partial_{r_w} T_b \right)
                \,\right]
\nonumber \\
\label{eq:beh.k3}
 k_3 &\defeq& - \left( \partial_M F_b \right)\,\left( \partial_H T_b \right)
         + \left( \partial_H F_b \right)\,\left( \partial_M T_b \right) \, .
\eqae
\seqe
The concrete forms of $k_i$ ($i = 1$, $2$, $3$) are obtained from the differentials of $T_b$ and $F_b$ shown in Sec.\ref{sec:beh.1}, and result in relations, 
\eqb
\label{eq:beh.k.rel}
 k_1 = M\,k_3 \quad,\quad k_2 = -H\,k_3 \, .
\eqe
Then Eq.\eref{eq:beh.pde.k} reduces to
\eqb
 M\,\pd{X_b}{M} - H\,\pd{X_b}{H} + r_w\,\pd{X_b}{\,r_w} = 2\,X_b \, .
\label{eq:beh.pde}
\eqe
This partial differential equation (PDE) is equivalent to the relation~\eref{eq:beh.euler.2} which is also equivalent to the relation~\eref{eq:beh.euler.1}. 
Therefore, if a solution $X_b$ of our PDE~\eref{eq:beh.pde} exists, then the $X_b$ satisfies the extensive scaling behavior $X_b \to \lambda^2\,X_b$ under the length size scaling~\eref{eq:ass.scaling}. 
Indeed, the general solution of PDE~\eref{eq:beh.pde} is expressed as
\eqb
 X_b(M,H,r_w) = r_w^2 \,\tilde{\psi}_b( M/r_w , H r_w ) \, ,
\label{eq:beh.Xb.sol}
\eqe
where $\tilde{\psi}_b(x,y)$ is an arbitrary function of two arguments. 
Obviously this $X_b$ is proportional to $r_w^2$, and satisfies the extensive scaling law under the length size scaling~\eref{eq:ass.scaling}. 
It is also obvious that the arbitrary functions $\Psi_b(\kappa_c r_w)$ in Eq.\eref{eq:ass.X.1} and $\Psi_b(H r_w)$ in Eq\eref{eq:ass.X.2} are consistent with $\tilde{\psi}_b(M/r_w , H r_w)$ in Eq.\eref{eq:beh.Xb.sol}, since $\kappa_c$ in Eq.\eref{eq:ass.kappa} is expressed as a function of $M/r_w$ and $H r_w$. 
Hence we find that the Euler relation~\eref{eq:beh.euler.2} is consistent with the assumption~4, which implies that the internal energy $E_b$ and also the free energy $F_b$ are defined well in our thermodynamics of BEH. 
The well-defined free energy guarantees the thermodynamic consistency. 
Now it has been checked that the minimal set of assumptions introduced in Sec.\ref{sec:ass} constructs the ``consistent'' thermodynamics for BEH.

\section{Cosmological event horizon}
\label{sec:ceh}

We examine whether the entropy-area law holds for ``consistent'' thermodynamics of our thermal equilibrium system for CEH.
Discussion in this section goes parallel to Sec.\ref{sec:beh}. 
However the integration constant in Euclidean action, which is determined with referring to de~Sitter canonical ensemble, enables us to find a reasonable evidence of the breakdown of entropy-area law for CEH.

\subsection{Temperature and free energy of CEH}
\label{sec:ceh.1}

By the assumption~3, the temperature $T_c$ of CEH is defined by Eq.\eref{eq:app.action.T} of Appendix~\ref{app:action}, which relates $T_c$ to the proper length in the imaginary time direction at the boundary $\partial D_c$,
\eqb
 T_c \defeq \left[\, \int_0^{\beta_c}\,\sqrt{f_w}\,d\tau \,\right]^{-1}
     = \frac{\kappa_c}{2\,\pi\,\sqrt{f_w}} \, ,
\label{eq:ceh.Tc}
\eqe
where $\beta_c$ is the imaginary time period~\eref{eq:action.period.ceh} and $f_w$ is in Eq.\eref{eq:action.fw}. 
This $T_c$ coincides with the Hawking temperature obtained by Gibbons and Hawking~\cite{ref:temperature}, and the factor $\sqrt{f_w}$ is the Tolman factor ~\cite{ref:tolman} which expresses the gravitational redshift affecting the Hawking radiation propagating from CEH to observer at $r_w$. 
Under the length size scaling~\eref{eq:ass.scaling}, this temperature satisfies the extensive scaling law, $T_c \to \lambda^{-1}\,T_c$.

As defined in assumption~2, the extensive state variable of system size for our thermal equilibrium system is the surface area of heat wall,
\eqb
 A \defeq 4\,\pi\,r_w^2 \, .
\eqe

By the assumption~3, the free energy $F_c$ of CEH in Eq.\eref{eq:ass.Fc} is defined by Eq.\eref{eq:app.action.F},
\eqb
 F_c(T_c,A,X_c) \defeq - T_c\,I_{Ec} = r_c\,\sqrt{f_w} + \frac{3\,M - r_c}{2\,\sqrt{f_w}} \, .
\label{eq:ceh.Fc}
\eqe
Under the length size scaling~\eref{eq:ass.scaling}, this free energy satisfies the scaling law of thermodynamic functions, $F_c \to \lambda\,F_c$. 
Furthermore, by the same discussion given in Sec.\ref{sec:beh.1}, it is also mathematically verified that the free energy $F_c$ must be a function of three independent variables, which verifies the working hypothesis~1.

Let us show some differentials for later use:
\seqb
\eqab
 \pd{T_c}{M} &=&
  \frac{1}{2 \pi r_c^2 f_w^{3/2}}
  \left[ \frac{r_c}{2 r_w} \left( 3 H^2 r_c^2 - 1 \right)
       - \frac{3 H^2 r_c^2 + 1}{3 H^2 r_c^2 - 1}\, f_w \right]
\label{eq-ceh.dTc/dM} \, , \\
 \pd{T_c}{H} &=&
  \frac{1}{4 \pi H r_c f_w^{3/2}}
  \left[ \left( 1 - \frac{2 M}{r_w} \right) \left( 3 H^2 r_c^2 -1 \right)
       - \frac{3 H^2 r_c^2 + 1}{3 H^2 r_c^2 - 1}\,\, \frac{2 M f_w}{r_c} \right]
\label{eq-ceh.dTc/dH} ,
\eqae
and
\eqab
\label{eq:ceh.dFc/dM}
 && \pd{F_c}{M} =
  \pd{\left[\,(r_c - r_w)\,\sqrt{f_w}\,\right]}{M}
  + \frac{1}{2 f_w^{3/2}}
    \left[ - \frac{r_c - 3 M}{r_w}
           + \frac{3 H^2 r_c^2 + 1}{3 H^2 r_c^2 - 1}\, f_w \right] \, , \\
\label{eq:ceh.dFc/dH}
 && \begin{aligned}
 \pd{F_c}{H} = \,\,& \pd{\left[\,(r_c - r_w)\,\sqrt{f_w}\,\right]}{H} \\
  &+ \frac{1}{2 f_w^{3/2}}
    \left[- \left( 2 r_w + r_c - 7 M - 2 H^2 r_w^3 \right) H^2 r_w^2
          + \frac{2 H r_c^3 f_w}{3 H^2 r_c^2 - 1} \right] \, ,
    \end{aligned}
\eqae
\seqe
where the differentials in Eqs.\eref{eq:ass.diff.rc} and~\eref{eq:ass.diff.kappac} are used. 
Then we get
\seqb
\label{eq:ceh.dFc.dTc}
\eqab
 \pd{F_c}{M} &=& \pd{\left[\,(r_c - r_w)\,\sqrt{f_w}\,\right]}{M} - \pi\,r_c^2\,\pd{T_c}{M}
\label{eq:ceh.dFc/dM.dTc/dM} \, , \\
 \pd{F_c}{H} &\not\propto& \pi\,r_c^2\,\pd{T_c}{H}
\label{eq:ceh.dFc/dH.dTc/dH} \, ,
\eqae
\seqe
where the definition of $r_c$, $f(r_c) = 0$, is used in the first relation. 
These relations are important to get a reasonable evidence of the breakdown of entropy-area law for CEH in next subsection.

Here one might naively expect that Eqs.\eref{eq:ceh.dFc/dM} and~\eref{eq:ceh.dFc/dH} would be obtained by replacing $r_b$ with $r_c$ in Eqs.\eref{eq:beh.dFb/dM} and~\eref{eq:beh.dFb/dH}, and also a relation $\partial_M F_c = - \pi r_c^2\, \partial_M T_c$ would be expected. 
However the first terms in the right-hand sides in Eqs.\eref{eq:ceh.dFc/dM}, ~\eref{eq:ceh.dFc/dH} and~\eref{eq:ceh.dFc/dM.dTc/dM} appear, because of the difference of integration constant in Euclidean action as seen in Eqs.\eref{eq:action.I0.beh} and~\eref{eq:action.I0.ceh}. 
The integration constant of $I_{Ec}$ can not be obtained by replacing $r_b$ with $r_c$ in that of $I_{Eb}$.

\subsection{Entropy of CEH}
\label{sec:ceh.2}

By the assumption~3, the entropy $S_c$ of CEH is defined as the thermodynamic conjugate variable to $T_c$,
\eqb
 S_c \defeq - \pd{F_c(T_c,A,X_c)}{T_c}
 = -\frac{(\partial_M F_c)\,(\partial_H X_c) - (\partial_H F_c)\,(\partial_M X_c)}
       {(\partial_M T_c)\,(\partial_H X_c) - (\partial_H T_c)\,(\partial_M X_c)} \, ,
\label{eq:ceh.Sc}
\eqe
where Eq.\eref{eq:app.formulas.2parameters} of Appendix~\ref{app:formulas} is used in the second equality. 
Under the length size scaling~\eref{eq:ass.scaling}, this entropy satisfies the extensive scaling law, $S_c \to \lambda^2\,S_c$. 
From this definition and the assumption~4 together with Eqs.\eref{eq:ceh.dFc.dTc}, we obtain an evidence of the breakdown of entropy-area law for CEH by the reductive absurdity as follows:

Assume that the entropy-area law holds for CEH, $S_c = \pi\,r_c^2$. 
Then Eq.\eref{eq:ceh.Sc} and $S_c = \pi\,r_c^2$ reduce to a PDE of $X_c$,
\seqb
\eqb
\label{eq:ceh.pde.1}
 J_M\,\pd{X_c}{M} + J_H\,\pd{X_c}{H} = 0 \, ,
\eqe
where  Eq.\eref{eq:ceh.dFc/dM.dTc/dM} is used, and
\eqb
 J_M \defeq \pd{F_c}{H} + \pi\,r_c^2\,\pd{T_c}{H} \quad,\quad
 J_H \defeq - \pd{\left[\,(r_c - r_w)\,\sqrt{f_w}\,\right]}{M} \, .
\eqe
\seqe
We find $J_M \not\equiv 0$ due to Eq.\eref{eq:ceh.dFc/dH.dTc/dH}, and $J_H \not\equiv 0$ due to $\partial_M r_c \not\equiv 0$ and $\partial_M f_w = -2/r_w$.

For the case of $X_c = r_w^2 \Psi_c(\kappa_b r_w)$ given in Eq.\eref{eq:ass.X.1} of assumption~4, the PDE~\eref{eq:ceh.pde.1} results in a contradiction as follows: 
Note that, because the surface gravities $\kappa_b$ and $\kappa_c$ are independent as functions of two variables $(M,H)$ due to non-zero \emph{Wronskian} $(\partial_M \kappa_b)\,(\partial_H \kappa_c) - (\partial_H \kappa_b)\,(\partial_M \kappa_c) \not\equiv 0$, the three quantities $(\kappa_b,\kappa_c,r_w)$ can be regarded as independent variables instead of $(M,H,r_w)$. 
Here, the transformation of independent variables between two pairs $(M,H,r_w)$ and $(\kappa_b,\kappa_c,r_w)$ is interpreted as the coordinate transformation in the state space of thermal equilibrium states of CEH. 
Then we find $\partial_{\kappa_c} X_c \equiv 0$ for $X_c = r_w^2 \Psi_c(\kappa_b r_w)$, and the PDE~\eref{eq:ceh.pde.1} reduces to 
\eqb
 \left( J_M\,\pd{\kappa_b}{M} + J_H\,\pd{\kappa_b}{H} \right)\,\pd{X_c}{\kappa_b} = 0 \, ,
\eqe
which gives $\partial_{\kappa_b} X_c \equiv 0$ due to $J_M\,\partial_M \kappa_b + J_H\,\partial_H \kappa_b \not\equiv 0$. 
On the other hand, the form of $X_c = r_w^2 \Psi_c(\kappa_b r_w)$ means $\partial_{\kappa_b} X_c \not\equiv 0$, since $\Psi_c$ is not constant as explained in assumption~4. 
Hence we find the PDE~\eref{eq:ceh.pde.1}, which is equivalent to the entropy-area law, contradicts Eq.\eref{eq:ass.X.1} of assumption~4.

Next, for the case of $X_c = r_w^2 \Psi_c(M/r_w)$ given in Eq.\eref{eq:ass.X.2}, the PDE~\eref{eq:ceh.pde.1} results in a contradiction as follows: 
With regarding the three quantities $(M,H,r_w)$ as independent variables, Eq.\eref{eq:ass.X.2} means $\partial_H X_c \equiv 0$ and the PDE~\eref{eq:ceh.pde.1} gives $\partial_M X_c \equiv 0$ due to $J_M \not\equiv 0$. 
On the other hand, the form of $X_c = r_w^2 \Psi_c(M/r_w)$ means $\partial_M X_c \not\equiv 0$, since $\Psi_c$ is not constant. 
Hence we find the PDE~\eref{eq:ceh.pde.1}, which is equivalent to the entropy-area law, contradicts Eq.\eref{eq:ass.X.2} of assumption~4.

\emph{The above discussions imply the breakdown of entropy-area law by the reductive absurdity under the minimal set of assumptions introduced in Sec.\ref{sec:ass}. 
Now it is concluded that we find a ``reasonable'' evidence of the breakdown of entropy-area law for CEH in SdS spacetime, where the ``reasonableness'' means that our discussion retains the ``thermodynamic consistency'' as shown in next subsection.}

\subsection{Thermodynamic consistency of CEH}
\label{sec:ceh.3}

The remaining part of Sec.\ref{sec:ceh} is for the ``thermodynamic consistency'' of our thermodynamics of CEH under the minimal set of assumptions. 
It is also verified that $X_c$ is an extensive variable and proportional to $r_w^2$. 
The discussion given to BEH in Sec.\ref{sec:beh.3} is applied to CEH.

By the assumption~3, the internal energy $E_c(S_c,A,X_c)$ of CEH, the surface pressure $\sigma_c$ at heat wall and the intensive variable $Y_c$ conjugate to $X_c$ are defined by
\eqab
\label{eq:ceh.Ec}
 E_c
 &\defeq& - \left.\pd{\ln Z_{cl}}{(1/T_c)}\right|_{A, X_c = \mbox{const.}}
  = \pd{(F_c/T_c)}{(1/T_c)}
  =  F_c + T_c\,S_c \, , \\
 \sigma_c &\defeq& - \pd{F_b(T_c,A,X_c)}{A} \nonumber \\
\label{eq:ceh.sigmac}
 &=& \mbox{Eq.\eref{eq:beh.sigmab} with replacing $(F_b , X_b)$ with $(F_c , X_c)$} \\
 Y_c
 &\defeq& \pd{ F_c(T_c , A , X_c)}{X_c} \nonumber \\
\label{eq:ceh.Yc}
 &=& \mbox{Eq.\eref{eq:beh.Yb} with replacing $(F_b , X_b)$ with $(F_c , X_c)$} \,,
\eqae
where the relation in Eq.\eref{eq:ceh.Ec}, $E_c(S_c,A,Y_c) = F_b(T_c,A,Y_c) + T_c\,S_c$, is regarded as the Legendre transformation, and Eqs.\eref{eq:app.formulas.3parameters} and~\eref{eq:app.formulas.2parameters} of Appendix~\ref{app:formulas} are used in the second equalities in $\sigma_c$ and $Y_c$. 
Under the length size scaling~\eref{eq:ass.scaling}, $E_c$ satisfies the scaling law of thermodynamic functions $E_c \to \lambda\,E_c$, and $\sigma_c$ and $Y_c$ satisfy the intensive scaling law $\sigma_c \to \lambda^{-1}\,\sigma_c$ and $Y_c \to \lambda^{-1}\,Y_c$. 
We find that, since the internal energy is a function of only extensive state variables, the state variable $X_c$ of BEH's gravitational effect on CEH should be an extensive variable. 
(The proportionality of $X_c$ to $r_w^2$ will also be shown mathematically at the end of this subsection.)

Then by these definitions of $E_c$, $\sigma_c$ and $Y_c$ together with the definition of $S_c$, we get the first law for CEH,
\eqb
 dE_c(S_c,A,Y_c) = T_c\,dS_c - \sigma_c\,dA + Y_c\,dX_c \, .
\eqe
Furthermore, the scaling behavior required in assumption~2 results in the same Euler relations as in Eqs.\eref{eq:beh.euler.1} and~\eref{eq:beh.euler.2},
\seqb
\eqab
\label{eq:ceh.euler.1}
 \lambda\, E_c(S_c \,,\, A \,,\, X_c) 
   &=& E_c(\lambda^2 S_c \,,\, \lambda^2 A \,,\, \lambda^2 X_c) \, , \\
\label{eq:ceh.euler.2}
 \frac{1}{2}\, E_b(T_c \,,\, A \,,\, X_c)
   &=& T_c\,S_c - \sigma_c\,A + Y_c\,X_c \, .
\eqae
\seqe
These two relations are mathematically equivalent by the Euler's theorem on the homogeneous expression.

By the Legendre transformation in Eq.\eref{eq:ceh.Ec} and the Euler relation~\eref{eq:ceh.euler.2}, we get a relation, $F_c = T_c\,S_c - 2\,\sigma_c\,A + 2\,Y_c\,X_c$. 
Then substituting $S_c$, $\sigma_c$ and $Y_c$ into this relation, we obtain a PDE of $X_c$,
\eqb
 l_1\,\pd{X_c}{M} + l_2\,\pd{X_c}{H} + r_w\,l_3\,\pd{X_c}{\,r_w}
 = 2\,l_3\,X_c \, ,
\label{eq:ceh.pde.l}
\eqe
where $l_i$ ($i = 1$, $2$, $3$) are defined formally by the same definitions of $k_i$ in Eqs.\eref{eq:beh.k1}, \eref{eq:beh.k2} and~\eref{eq:beh.k3} by replacing $(F_b\,,\,T_b)$ with $(F_c\,,\,T_c)$.
Using the differentials of $T_c$ and $F_c$ shown in Sec.\ref{sec:ceh.1}, we find $l_1 = M\,l_3$ and $l_2 = -H\,l_3$ which is the same with Eq.\eref{eq:beh.k.rel}.
Then our PDE~\eref{eq:ceh.pde.l} reduces to the same PDE in Eq.\eref{eq:beh.pde}, and its general solution is
\eqb
 X_c(M,H,r_w) = r_w^2 \,\tilde{\psi}_c( M/r_w , H r_w ) \, ,
\label{eq:ceh.Xc.sol}
\eqe
where $\tilde{\psi}_c(x,y)$ is an arbitrary function of two arguments. 
Obviously this $X_c$ is proportional to $r_w^2$, and satisfies the extensive scaling law under the length size scaling~\eref{eq:ass.scaling}. 
It is also obvious that the arbitrary functions $\Psi_c(\kappa_b r_w)$ in Eq.\eref{eq:ass.X.1} and $\Psi_c(M/r_w)$ in Eq\eref{eq:ass.X.2} are consistent with $\tilde{\psi}_c(M/r_w , H r_w)$ in Eq.\eref{eq:ceh.Xc.sol}, since $\kappa_b$ in Eq.\eref{eq:ass.kappa} is expressed as a function of $M/r_w$ and $H r_w$. 
Hence we find that the Euler relation~\eref{eq:ceh.euler.2} is consistent with the assumption~4, which implies that the internal energy $E_c$ and also the free energy $F_c$ are defined well in our thermodynamics of CEH. 
The well-defined free energy guarantees the thermodynamic consistency. 
Now it has been checked that the minimal set of assumptions introduced in Sec.\ref{sec:ass} constructs the ``consistent'' thermodynamics for CEH.

\section{Summary and discussions}
\label{sec:sd}

\subsection{Summary and comments on BEH's entropy}
\label{sec:sd.1}

To research whether the thermal equilibrium is the necessary and sufficient condition to ensure the entropy-area law, we have carefully constructed \emph{two thermal equilibrium systems} individually for BEH and CEH in SdS spacetime, and the ``consistent thermodynamics'' have been obtained for BEH and CEH under the minimal set of assumptions. 
The need of those assumptions was discussed in Sec.\ref{sec:ass.preliminary}. 
In the construction of the two thermal equilibrium systems, the role of cosmological constant in the consistent thermodynamics has also been pointed out in the working hypothesis~1. 
In our analysis, Euclidean action method was used with referring to Schwarzschild and de~Sitter canonical ensembles to determine the integration constants (subtraction terms). 
As a result, we have found a reasonable evidence for the breakdown of entropy-area law for CEH as shown in Sec.\ref{sec:ceh.2}, while the validity of the law for BEH could not be judged but the key issue on BEH's entropy has been clarified in Sec.\ref{sec:beh.2}. 
If the breakdown of the law for BEH is verified, then it means that, as summarized in fourth paragraph in Sec.\ref{sec:intro}, the thermal equilibrium of each horizon in multi-horizon spacetime is simply a necessary condition of the entropy-area law, and the necessary and sufficient condition of the law is the thermal equilibrium of the total system composed of several horizons in which the net energy flow among horizons disappears.

The analysis in this paper is exact for parameter range, $0 < \sqrt{27}\,M H <1$ and $r_b < r_w < r_c$, where the first inequality ensures that the BEH and CEH is non-degenerate, $r_b < r_c$. 
It is obvious that our discussion is true of the near Nariai case (near extremal case of SdS spacetime), $r_b \simeq r_c$ ($\,\Leftrightarrow\,\sqrt{27}\,M H \simeq 1$)~\footnote{The metric of extremal SdS spacetime was found by Nariai~\cite{ref:nariai}, independently of the non-extreme SdS metric by Kottler~\cite{ref:kottler}.}. 
However note that the temperatures of horizons are equal at the exact Nariai case, $T_b = T_c$ at $r_b = r_c$. 
Then, Appendix~\ref{app:nariai} analyzes how the near Nariai case affects the entropy-area law.

Turn our attention to the entropy of BEH in general case (not at Nariai limit). 
Some comments which suggest the breakdown of entropy-area law for BEH may be possible. 
Let us try to give three comments: 
For the first, recall the meaning of state variable $X_b$, which expresses the thermodynamic effect on BEH due to the external gravitational field produced by CEH. 
Furthermore it is worth pointing out that the CEH temperature $T_c$ depends on $\kappa_c$ which has dependence on $(M\,,\,H)$, not on $H$ solely. 
Then, in the assumption~4, it may be natural that the quantity $\kappa_c r_w$, not $H r_w$, is the characteristic variable of CEH's gravity and the extensive variable $X_b$ is expressed by $X_b = r_w^2 \Psi_b(\kappa_c r_w)$ as required in Eq.\eref{eq:ass.X.1}. 
If this is true, then the breakdown of entropy-area law for BEH is concluded as explained in Sec.\ref{sec:beh.2}.

Next, recall that, in Sec.\ref{sec:ass.assumption1}, our thermal equilibrium systems for BEH and CEH are compared qualitatively with the magnetized gas. 
By the differential of free energy $F_{\rm gas}(T_{\rm gas} , V_{\rm gas} , \vec{H}_{\rm gad})$, the entropy $S_{\rm gas}$ and pressure $P_{\rm gas}$ of the gas are defined by (see for example \S52, 59 and~60 in Landau and Lifshitz~\cite{ref:ll}),
\eqb
 S_{\rm gas} \defeq
  -\pd{F_{\rm gas}(T_{\rm gas} , V_{\rm gas} , \vec{H}_{\rm ex})}{T_{\rm gas}} \quad,\quad
 P_{\rm gas} \defeq
  -\pd{F_{\rm gas}(T_{\rm gas} , V_{\rm gas} , \vec{H}_{\rm ex})}{V_{\rm gas}} \, .
\eqe
This implies that all state variables of the gas depend on the external field $\vec{H}_{\rm ex}$. 
Therefore, the entropy of the gas under the influence of external magnetic field deviates from the entropy without external magnetic field. 
Hence, for our thermal equilibrium system for BEH, it may be naturally expected that BEH's entropy under the influence of external gravitational field of CEH does not satisfy the entropy-area law which holds for BEH in single-horizon spacetimes.

For the third comment, we note that, while the CEH temperature $T_c$ is obtained from BEH temperature $T_b$ by the simple replacement of $(r_b\,,\,\kappa_b)$ with $(r_c\,,\,\kappa_c)$, the CEH free energy $F_c$ can not be obtained from BEH free energy $F_b$ by such a simple replacement. 
This ``asymmetry'' of $F_b$ and $F_c$ is due to the asymmetry of integration constant of BEH's Euclidean action~\eref{eq:action.I0.beh} and that of CEH's one~\eref{eq:action.I0.ceh}. 
Then it is naively expected that the coefficients $l_i$ of PDE~\eref{eq:ceh.pde.l} do not satisfy the same relation~\eref{eq:beh.k.rel} as $k_i$. 
However we find at Eq.\eref{eq:ceh.pde.l} that the relation~\eref{eq:beh.k.rel} holds for both coefficients $k_i$ and $l_i$, and the same expression of general solutions of $X_b$ and $X_c$ are obtained as shown in Eqs.\eref{eq:beh.Xb.sol} and~\eref{eq:ceh.Xc.sol}. 
This may imply that the same consistent structure of thermodynamics holds for BEH and CEH even though the forms of free energies are asymmetric. 
If this implication is true, then, since the entropy-area law breaks down for CEH, the law for BEH may also break down.

The above three comments have no rigorous verification. 
Those verifications are left as future tasks, but the key issue is already clarified in Sec.\ref{sec:beh.2}.

\subsection{Discussions}

Let us try to make two discussions. 
One of them is on the quantum statistics of underlying quantum gravity, and the other is on SdS black hole evaporation as a non-equilibrium process. 
These are independent of each other.

For the first, we discuss about the quantum statistics: 
The analysis in the main text of this paper is based on the Euclidean action method. 
This is equivalent to assume that the basic principle of statistical mechanics of ordinary laboratory systems works well in calculating the partition function of the canonical ensemble for our two thermal equilibrium systems. 
Here let us emphasize that the basic principle of statistical mechanics of ordinary laboratory systems is the \emph{principle of equal a priori probabilities}~\cite{ref:sm,ref:ll}~\footnote{
When this principle is applied to the micro-canonical ensemble, the Boltzmann's relation $S = \ln W$ is obtained, where $S$ and $W$ are respectively the entropy and the number of states. And when this principle is applied to the canonical ensemble, the free energy is obtained by the relation in Eq.\eref{eq:app.action.F}.
}.
Hence, if the analysis in this paper and the comments in previous subsection are true (to imply the breakdown of entropy-area law for BEH), then it suggests that the principle of equal a priori probabilities results in the breakdown of entropy-area law for multi-horizon spacetimes in which horizon temperatures are not equal and a net energy flow among horizons exits. 
In other words, if the statistics of micro-states of quantum gravity obeys the principle of equal a priori probabilities, then the entropy-area law breaks down for the multi-horizon spacetimes.

Then what will be suggested if we adopt the other point of view? 
Let us dare to give priority to the entropy-area law, and assume that the entropy-area law holds for the two thermal equilibrium systems constructed in the assumption~1. 
Under this assumption, the discussion in previous paragraph implies that the principle of equal a priori probabilities and the Euclidean action method is not suitable to the quantum statistics of gravity in the multi-horizon spacetimes. 
In this case, the underlying quantum gravity should be formulated to yield the special statistic property of micro-states of gravity, which comes to obey the principle of equal a priori probability in the case of single-horizon limit~\cite{ref:euclidean.sch,ref:euclidean.kn,ref:euclidean.sads,ref:euclidean.ds}.

Next turn our discussion to the second one, which is completely separated from the above discussion of quantum statistics. 
The second discussion is on SdS black hole evaporation process: 
Hereafter the heat wall introduced in the assumption~1 is \emph{removed}. 
Let us note the inequality $T_b > T_c$ due to Eq.\eref{eq:ass.kappa.rel}, which means the existence of a net energy flow from BEH to CEH due to the exchange of Hawking radiation emitted by two horizons. 
This means that, as mentioned in Sec.\ref{sec:intro}, the region~I in SdS spacetime is in a \emph{non-equilibrium state}, and the SdS spacetime evolves in time due to the energy flow. 
This time evolution is the SdS black hole evaporation process. 
Here note that Hawking temperature is usually much lower than the energies $E_b$ and $E_c$ when the horizon is not quantum but classical size~\cite{ref:hr}. 
Then the evolution of BEH and CEH during the SdS black hole evaporation can be described by the so-called \emph{quasi-static process}, in which thermodynamic states of BEH and CEH at each instant of the evolution can be approximated well by thermal equilibrium states. 
(Thermodynamic state of BEH evolves on a path in the state space of thermal equilibrium states. 
Also the thermodynamic state of CEH do the same, but the path in state space on which CEH evolves is different from that of BEH.) 
This implies that the matter field of Hawking radiation is responsible for the non-equilibrium nature of SdS spacetime. 
Because the matter field is in a non-equilibrium state, the total system composed of SdS spacetime and the matter field of Hawking radiation is in a non-equilibrium state, even when the horizons are individually in equilibrium states~\footnote{
The matter field of Hawking radiation is neglected in the main context of this paper, because its energy scale is negligible for classical size horizons~\cite{ref:hr}. 
However, when we proceed to the research on the non-equilibrium nature of SdS spacetime and its time evolution, it is necessary to consider the matter field of Hawking radiation which is responsible for the non-equilibrium nature of SdS evaporation process.
}. 
If we can formulate a general non-equilibrium thermodynamics for arbitrary matter fields which are enclosed by two thermal bodies of different temperatures, then a non-equilibrium SdS thermodynamics may be obtained by applying the non-equilibrium thermodynamics to matter fields of Hawking radiation.

For non-self-interacting matters, a two-temperature non-equilibrium thermodynamics has already been constructed~\cite{ref:noneq.rad}. 
Therefore, under the assumption that the matter field of Hawking radiation is non-self-interacting (e.g. a minimal coupling massless scaler field~$\phi$ satisfying $\Box \phi = 0$), the non-equilibrium evolution process of SdS spacetime may be described by using the non-equilibrium thermodynamics of non-self-interacting matters~\cite{ref:noneq.rad}. 
Indeed, the non-equilibrium thermodynamics of non-self-interacting matters has already been applied to the evaporation process of Schwarzschild black hole~\cite{ref:evapo}, and has revealed the detail of evaporation process as a non-equilibrium process and verified the so-called generalized second law for the evaporation process~\cite{ref:gsl}. 
However the non-equilibrium thermodynamics of non-self-interacting matters requires to know \emph{a priori} the equilibrium state variables of thermal bodies among which the non-equilibrium matter field is enclosed. 
Therefore, before applying the non-equilibrium thermodynamics~\cite{ref:noneq.rad} to SdS spacetime, we have to specify the state variables $X_b$ and $X_c$.

Finally for self-interacting matters, under the condition that its non-equilibrium nature is not so strong, some non-equilibrium thermodynamics have already been constructed~\cite{ref:noneq}. 
Therefore, under the assumption that the strength of non-equilibrium nature is not so strong (e.g. not so large temperature difference), the non-equilibrium evolution process of SdS spacetime may be described by using an appropriate non-equilibrium thermodynamics~\cite{ref:noneq}. 
Concerning self-interacting matters, as far as the author knows, no one has been applied the theories~\cite{ref:noneq} even to Schwarzschild black hole evaporation process.

\section*{Acknowledgments}

I'd like to express my gratitude to my colleague Hiroshi Harashina, the experimental physicist researching superconductivity and magnetic substances, for his continuous communication about thermodynamics and statistical mechanics of ordinary laboratory systems. 
Also I thank my colleague Toshihide Futamura, the mathematician, for his help to notice useful formulas displayed in Appendix~\ref{app:formulas}. 
This work is supported by the Grant-in-Aid for Scientific Research Fund of the Ministry of Education, Culture, Sports, Science and Technology, Japan (Young Scientists (B) 19740149).

\appendix
\section{Definitions and formulas in Euclidean action method}
\label{app:action}

Thermal field theory~\cite{ref:tft} is a statistical mechanics for quantum matter fields in thermal equilibrium in flat spacetime, which is well verified experimentally for laboratory systems. 
It is generalized by Gibbons and Hawking~\cite{ref:euclidean} to give a postulated formulas of temperature and free energy for thermal equilibrium systems of quantum gravity. 
This is called the Euclidean action method. 
Instead of the original works by Gibbons and Hawking~\cite{ref:euclidean}, Appendix~A in previous work~\cite{ref:euclidean.ds} may be sufficient for a review of Euclidean action method. 
This appendix lists only the definitions and formulas of Euclidean action method used in the main text of this paper.

The Wick rotation is the transformation of time coordinate $t$ of Lorentzian spacetime to the imaginary value $-i \tau$ in the clockwise direction,
\eqb
 \text{Wick rotation} : t \to - i\, \tau \, .
\eqe
The Euclidean action of spacetime is defined by
\eqb
 I_E[g_E] \defeq
 i\times \mbox{$I_L[g]$ with Wick rotation $t \to -i\,\tau$} \, ,
\label{eq:app.action.IE.curved}
\eqe
where $g_E$ is the Euclidean metric of signature $(+ + + +)$ obtained from the Lorentzian metric $g$ by the Wick rotation and $I_L[g]$ is the Lorentzian action of spacetime. 
Here the signature of $g$ is $(- + + +)$.

In the Euclidean action method, $I_E[g_E]$ corresponds to the partition function $Z_{cl}$ of thermal equilibrium states of the spacetime under consideration:
\eqb
 \ln Z_{cl} \defeq I_E[g_{E\,cl}] \, ,
\label{eq:app.action.Zcl}
\eqe
where the lower suffix $_{cl}$ means that the metric $g_{E\,cl}$ in the argument of $I_E$ is the solution of classical Einstein equation.

The equilibrium temperature $T$ is defined by the proper length in imaginary time direction in the Euclidean space,
\eqb
 T \defeq \left[\, \int_0^{\beta}\,\sqrt{g_{E\,\tau \tau}}\,d\tau \,\right]^{-1} \, ,
\label{eq:app.action.T}
\eqe
where $\beta$ is the period of imaginary time $\tau$ determined by the regularity requirement of Euclidean space at the event horizon. 
Since the metric component $g_{E\,\tau \tau}$ is a function of spacetime coordinates, the integral in Eq.\eref{eq:app.action.T} becomes a function of spatial coordinates. 
Therefore it is important to specify where the temperature is defined. 
Here note that the Euclidean action method is for the canonical ensemble. 
This implies the existence of a heat bath whose temperature coincides with the temperature of the system under consideration, since the system is in a thermal equilibrium with the heat bath. 
Therefore it is reasonable to evaluate $g_{E\,\tau \tau}$ at the contact surface between the system and the heat bath. 
The contact surface is the boundary of the spacetime region. 
Hence the temperature $T$ should be evaluated at the spacetime boundary.

Then the free energy of thermal equilibrium states of the spacetime under consideration is defined by
\eqb
 F \defeq -T\,\ln Z_{cl} = - T\,I_E[g_{E\,cl}] \, ,
\label{eq:app.action.F}
\eqe
where $T$ is defined by Eq.\eref{eq:app.action.T} with replacing $g_E$ by $g_{E\,cl}$.

\section{Key points of Schwarzschild thermodynamics}
\label{app:key}

From the following key points of Schwarzschild thermodynamics, we can learn the minimal set of assumptions of the BEH's and CEH's thermodynamics. 
Indeed, thermodynamically consistent de~Sitter canonical ensemble has already been constructed with referring to the following key points in previous paper~\cite{ref:euclidean.ds}. 
There are three key points which should be shown here. 
The first one is the zeroth law which describes the existence of thermal equilibrium states:
\begin{key}[Zeroth law of black hole] 
Place a black hole in a spherical cavity as shown in Fig.\ref{fig:1} and also the observer at the surface of the heat bath. 
Through the Hawking radiation by black hole and the black body radiation by heat bath, the black hole interacts with the heat bath. 
Then, by appropriately adjusting the temperature of heat bath, the combined system of black hole and heat bath settles down to a thermal equilibrium state. 
This equilibrium state of black hole under the contact with heat bath is described by the canonical ensemble. 
Using the canonical ensemble, the equilibrium state variables of black hole are defined by the quantities measured at the surface of heat bath where the observer is. 
Then the ``thermodynamically consistent'' black hole thermodynamics is obtained as summarized below.
\end{key}

\begin{figure}[t]
 \begin{center}
 \includegraphics[height=30mm]{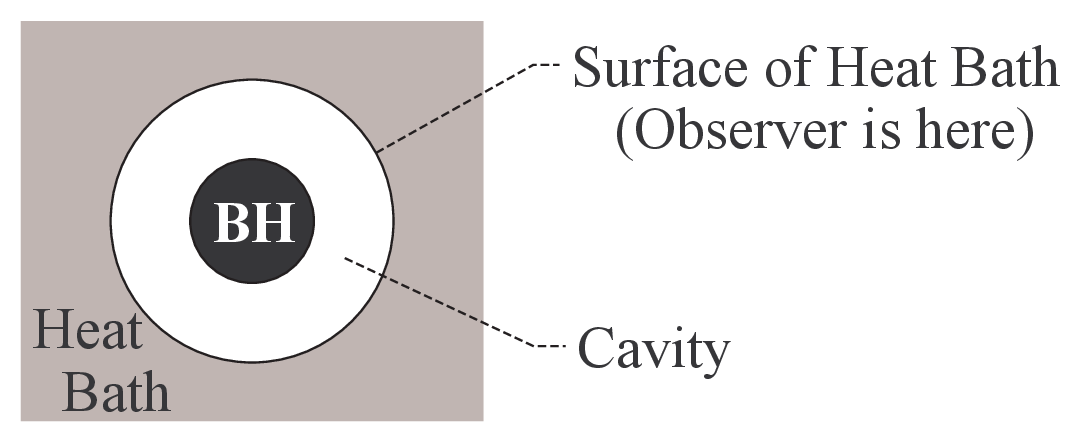}
 \end{center}
\caption{Schematic image of thermal equilibrium of black hole with heat bath. This is described by the canonical ensemble. State variables of black hole are defined at the surface of heat bath. With those state variables, the consistent thermodynamic formulation is realized using the Euclidean action method~\cite{ref:euclidean.sch}.}
\label{fig:3}
\end{figure}

The second key point is the difference of black hole thermodynamics from thermodynamics of ordinary laboratory systems: 
\begin{key}[Peculiar scaling law of black hole] 
Extensive and intensive state variables of black hole show a peculiar scaling law: 
When a length size $L$ (e.g. event horizon radius) is scaled as $L \to \lambda\,L$ with an arbitrary scaling rate $\lambda\,(> 0)$, then the extensive variables $X$ (e.g. entropy) and intensive variables $Y$ (e.g. temperature) are scaled as $X \to \lambda^2\,X$ and $Y \to \lambda^{-1}\,Y$, while the thermodynamic functions $\Phi$ (e.g. free energy) are scaled as $\Phi \to \lambda\,\Phi$. 
This implies that, because the system size is extensive, the thermodynamic system size of equilibrium system constructed in the key point~1 should have the areal dimension. 
Indeed the surface area of heat bath, $4 \pi r_w^2$, behaves as the consistent extensive variable of system size, where $r_w$ is the radius of the surface of heat bath. 
\end{key}

Here recall that, in thermodynamics of ordinary laboratory systems, the intensive variables remain un-scaled under the scaling of system size, the extensive variables scales as the volume (for the system of three spatial dimensions), and the thermodynamic functions are the members of extensive variables. 
However, as noted in the key point~2, the Schwarzschild thermodynamics has the peculiar scaling law of state variables. 
Although the scaling law differs from that in thermodynamics of ordinary laboratory systems, the peculiar scaling law in Schwarzschild thermodynamics (and also de~Sitter thermodynamics) retains the thermodynamic consistency as noted in the next key point.

The third key point is the similarity of black hole thermodynamics with thermodynamics of ordinary laboratory systems:
\begin{key}[Euclidean action method and thermodynamic consistency] 
The free energy $F_{\rm BH}$ of black hole is given by the Euclidean action method, where the integration constant (the so-called subtraction term) of the action integral is determined with referring to flat spacetime. 
The action integral is evaluated in the region denoted by $2 M < r < r_w$ which is in thermal equilibrium as noted in the key point~1, where $M$ is the mass parameter.
Then, as for the ordinary thermodynamics, this free energy is expressed as a function of two independent state variables, temperature and system size;
\eqb
 F_{\rm BH}(T_{\rm BH} , 4 \pi r_w^2) \,,
\eqe
where the intensive variable $T_{\rm BH} \defeq \left(8 \pi M \sqrt{1-2 M/r_w}\right)^{-1}$ is the Hawking temperature measured by the observer at $r_w$ and the factor $\sqrt{1-2 M/r_w}$ is the so-called Tolman factor~\cite{ref:tolman} which expresses the gravitational redshift affecting the Hawking radiation propagating from the black hole horizon to the observer. 
This Hawking temperature is obtained by Eq.\eref{eq:app.action.T} of Appendix~\ref{app:action}. 
(The temperature of heat bath should be adjusted to be $T_{\rm BH}$ in the key point~1.) 
In order to let $T_{\rm BH}$ and $4 \pi r_w^2$ be independent variables in $F_{\rm BH}$, the mass parameter $M$ and the heat bath radius $r_w$ are regarded as two independent variables. 
Then the thermodynamic consistency holds as follows: 
The entropy $S_{\rm BH}$ and the ``surface'' pressure $\sigma_{\rm BH}$ are defined by
\eqb
\label{eq:intro.S.sigma_BH}
 S_{\rm BH} \defeq -\pd{F_{\rm BH}(T_{\rm BH} , 4 \pi r_w^2)}{T_{\rm BH}}
                     = \frac{A_{\rm BH}}{4} \quad,\quad
 \sigma_{\rm BH} \defeq -\pd{F_{\rm BH}(T_{\rm BH} , 4 \pi r_w^2)}{ (4 \pi r_w^2)} \, ,
\eqe
where $\sigma_{\rm BH}$ has the dimension of force par unit area. 
(See Appendix~B in previous paper~\cite{ref:euclidean.ds} for a detail explanation of the thermodynamic meaning of $\sigma_{\rm BH}$.) 
These differential relations among the free energy, entropy and surface pressure are the same as obtained in thermodynamics of ordinary laboratory systems. 
Furthermore, as for the ordinary thermodynamics, the internal energy and the other thermodynamic functions are defined by the Legendre transformation of the free energy; for example the internal energy $E_{\rm BH}$ is
\eqb
 E_{\rm BH}(S_{\rm BH},4 \pi r_w^2) \defeq F_{\rm BH} + T_{\rm BH}\,S_{\rm BH} \,.
\eqe
The enthalpy, Gibbs energy and so on are also defined by the Legendre transformation. 
Then the differential relations among those thermodynamic functions and the other state variables also hold, for example $T_{\rm BH} \equiv \partial E_{\rm BH}/\partial S_{\rm BH}$. 
Furthermore, with the state variables obtained above, we can check that the first, second and third laws of thermodynamics hold for black holes.
\end{key}

The above three key points can hold also, at least, for the other single-horizon spacetimes, and the consistent thermodynamics has already been established for those single-horizon spacetimes~\cite{ref:euclidean.sch,ref:euclidean.kn,ref:euclidean.ds}.

Here let us comment on the heat bath introduced in the key point~1. 
In York's consistent black hole thermodynamics~\cite{ref:euclidean.sch}, the heat bath is essential to establish the thermodynamic consistency in the canonical ensemble as explained in the following: 
Generally in thermodynamics, as noted in the key point~3, thermodynamic functions are defined as a function of two independent state variables. 
Especially the free energy should be expressed as a function of the temperature and the extensive state variable which represents the system size. 
To satisfy such thermodynamic requirement, the introduction of heat bath gives us two independent variables; the mass parameter $M$ and the radius of heat bath $r_w$. 
These two independent variables makes it possible to define the temperature $T_{\rm BH}$ and the surface area $4 \pi r_w^2$ as the two independent state variables of free energy $F_{\rm BH}$. 
Therefore the heat bath is necessary to establish manifestly the thermodynamic consistency.

\section{Useful differential formulas}
\label{app:formulas}

\subsection{First case}

Let $f$ be a function of $\alpha_1$, $\alpha_2$ and $\alpha_3$, $f = f(\alpha_1,\alpha_2,\alpha_3)$.
And consider the case that $\alpha_i$ ($i = 1$,~$2$,~$3$) are also functions of $y_1$, $y_2$ and $y_3$, $\alpha_i = \alpha_i(y_j)$ ($j = 1$,~$2$,~$3$). 
Then define $f(y_1,y_2,y_3) \defeq f(\,\alpha_i(y_j)\,)$. 
Let us aim to express the partial derivatives $\partial f(\alpha_1,\alpha_2,\alpha_3)/\partial \alpha_i$ by the derivatives with respect to $y_j$. 
Standard differential calculus gives, $\partial_{y_j}f = \sum_{i=1}^3 (\partial_{\alpha_i}f) \, (\partial_{y_j}\alpha_i)$, which is expressed in vector form as
\eqb
 \begin{pmatrix}
   \partial_{y_1}f \cr
   \partial_{y_2}f \cr
   \partial_{y_3}f
 \end{pmatrix}
 =
 P\, \begin{pmatrix}
       \partial_{\alpha_1}f \cr
       \partial_{\alpha_2}f \cr
       \partial_{\alpha_3}f
      \end{pmatrix}
 \quad,\quad
 P \defeq
 \begin{pmatrix}
   \partial_{y_1}\alpha_1 & \partial_{y_1}\alpha_2 & \partial_{y_1}\alpha_3 \cr
   \partial_{y_2}\alpha_1 & \partial_{y_2}\alpha_2 & \partial_{y_2}\alpha_3 \cr
   \partial_{y_3}\alpha_1 & \partial_{y_3}\alpha_2 & \partial_{y_3}\alpha_3
 \end{pmatrix}
 \, .
\label{eq:app.formulas.vec}
\eqe
Then, when $\det P \neq 0$, we obtain
\eqb
 \begin{pmatrix}
   \partial_{\alpha_1}f \cr
   \partial_{\alpha_2}f \cr
   \partial_{\alpha_3}f
 \end{pmatrix}
 =
 P^{-1}\,
 \begin{pmatrix}
   \partial_{y_1}f \cr
   \partial_{y_2}f \cr
   \partial_{y_3}f
 \end{pmatrix}    \, .
\label{eq:app.formulas.3parameters}
\eqe
In section~\ref{sec:beh}, by setting $f = F_b$, $(\alpha_1,\alpha_2,\alpha_3) = (T_b,A,X_b)$ and $(y_1,y_2,y_3) = (M,H,r_w)$, we can obtain the conjugate state variables to $(T_b,A,X_b)$ as combinations of partial derivatives with respect to $(M,H,r_w)$. 
The same is also applied in section~\ref{sec:ceh}.

\subsection{Second case}

Use the same definitions with previous subsection. 
If $f$ has no $\alpha_3$-dependence ($f = f(\alpha_1,\alpha_2)$\,) and $\alpha_i$ has no $y_3$-dependence ($\alpha_i = \alpha_i(y_1,y_2)$, $i = 1$, $2$), then Eq.\eref{eq:app.formulas.3parameters} reduces to
\eqb
 \pd{f(\alpha_1,\alpha_2)}{\alpha_1}
 = \frac{(\partial_{y_1} f)\,(\partial_{y_2} \alpha_2)
         - (\partial_{y_2} f)\,(\partial_{y_1} \alpha_2)}
        {(\partial_{y_1} \alpha_1)\,(\partial_{y_2} \alpha_2)
         - (\partial_{y_2} \alpha_1)\,(\partial_{y_1} \alpha_2)}
\label{eq:app.formulas.2parameters} \, ,
\eqe
and a similar formula given by exchanging $\alpha_1$ and $\alpha_2$.

\subsection{Third case}
\label{app-formulas.third}

Use the same definitions with previous subsection. 
If $f$ has no $\alpha_3$-dependence ($f = f(\alpha_1,\alpha_2)$\,) and $\alpha_2$ has no $y_2$- and $y_3$-dependence ($\alpha_2 = \alpha_2(y_1)$\,) while $\alpha_1$ depends on all of $y_j$ ($\alpha_1 = \alpha_1(y_1,y_2,y_3)$\,), then Eq.\eref{eq:app.formulas.vec} reduces to
\eqab
 \begin{pmatrix}
   \partial_{y_1}f \cr
   \partial_{y_2}f \cr
   \partial_{y_3}f
 \end{pmatrix}
 =
 \begin{pmatrix}
   \partial_{y_1}\alpha_1 & \partial_{y_1}\alpha_2 \cr
   \partial_{y_2}\alpha_1 & 0 \cr
   \partial_{y_3}\alpha_1 & 0
 \end{pmatrix}\,
 \begin{pmatrix}
   \partial_{\alpha_1}f \cr
   \partial_{\alpha_2}f
 \end{pmatrix} \, .
\eqae
This gives
\eqab
 \pd{f(\alpha_1,\alpha_2)}{\alpha_1}
 &=& \frac{\partial_{y_2}f}{\partial_{y_2}\alpha_1}
     = \frac{\partial_{y_3}f}{\partial_{y_3}\alpha_1}
\label{eq:app.formulas.3rd} \\
 \pd{f(\alpha_1,\alpha_2)}{\alpha_2}
 &=& \frac{(\partial_{y_1} f)\,(\partial_{y_2} \alpha_1)
            - (\partial_{y_2} f)\,(\partial_{y_1} \alpha_1)}
          {(\partial_{y_1} \alpha_2)\,(\partial_{y_2} \alpha_1)} \\
 &=&  \frac{(\partial_{y_1} f)\,(\partial_{y_3} \alpha_1)
            - (\partial_{y_3} f)\,(\partial_{y_1} \alpha_1)}
          {(\partial_{y_1} \alpha_2)\,(\partial_{y_3} \alpha_1)}
\nonumber \, .
\eqae
Furthermore, if $f$ depends only on $\alpha_1$ ($f = f(\alpha_1)$\,), then Eq.\eref{eq:app.formulas.3rd} reduces to
\eqb
 \pd{f(\alpha_1)}{\alpha_1}
 = \frac{\partial_{y_1}f}{\partial_{y_1}\alpha_1}
 = \frac{\partial_{y_2}f}{\partial_{y_2}\alpha_1}
 = \frac{\partial_{y_3}f}{\partial_{y_3}\alpha_1}
\label{eq:app.formulas.3rd.2} \, .
\eqe

\section{Near Nariai case}
\label{app:nariai}

This appendix analyzes the near Nariai case (near extremal case of SdS spacetime) of our two thermal equilibrium systems of BEH and CEH. 
Note that, in the exact Nariai case, the two horizons degenerate and our two thermal equilibrium systems of horizons disappear. 
Hence we consider the near Nariai case as a perturbation of the exact Nariai case. 
We introduce two independent small parameters, $\delta_w$ and $\delta_{bc}$, defined by 
\eqb
 r_w \defeqr 3M + \delta_w \quad,\quad
 r_c \defeqr r_b + \delta_{bc} \,,
\eqe
where we require $\delta_{bc} \ll M$ which means the near Nariai case.
The parameter $\delta_w$ controls the position of the heat wall, and satisfies, $-(3 M - r_b) < \delta_w < r_c - 3 M$.

In the following, we expand the temperatures and free energies of our two thermal equilibrium systems by the small parameters $\delta_w$ and $\delta_{bc}$, in which the 0-th order values are of the Nariai limit $\delta_{bc} \to 0$ and $\delta_w \to 0$. 
We measure the size of the exact Nariai spacetime by the mass parameter $M$. 
Then, it is useful to introduce the following supplemental small parameters, $\delta_H$ and $\delta_\alpha$, defined by
\eqb
 H \defeqr \frac{1}{\sqrt{27}\,M} - \delta_H \quad,\quad
 \alpha \defeqr \frac{\pi}{2} - \delta_\alpha \,,
\eqe
where $\alpha$ is given by $\sin\alpha = \sqrt{27} M H$. 
One of three parameters $(\,\delta_{bc}\,,\,\delta_H\,,\,\delta_\alpha\,)$ is independent. 
By the definition of $\alpha$, $\sin\alpha = \sqrt{27} M H$, and Eq.\eref{eq:ass.r}, we obtain
\eqb
 2 \sqrt{27}\,M\,\delta_H = \delta_\alpha^2 + O(\delta_\alpha^4) \quad,\quad
 \delta_{bc} = 2 \sqrt{3}\,M\,\delta_\alpha\,
                 \left[\, 1 + \frac{1}{3}\,\delta_\alpha^2 + O(\delta_\alpha^4) \,\right] \,.
\eqe
Furthermore, from Eq.\eref{eq:action.fw}, we obtain
\eqb
 f_w = \frac{\delta_{bc}^2}{36 M^2}\,
       \left[\, 1 - 4 \left(\frac{\delta_w}{\delta_{bc}}\right)^2 \,\right]
       + O(\delta_{bc}^4) 
       + O(\delta_{bc}^2\,\delta_w) \,.
\eqe
The requirement $f_w > 0$  means $\bigl|\delta_w/\delta_{bc}\bigr| < 1/2$.

From the above, we obtain the near Nariai value of temperatures~\eref{eq:beh.Tb} and~\eref{eq:ceh.Tc},
\seqb
\eqb
 T_b = T_N\,\left[\, 1 + \frac{\delta_{bc}}{9 M} + O(\delta_{bc}^2) \,\right] \quad,\quad
 T_c = T_N\,\left[\, 1 - \frac{\delta_{bc}}{9 M} + O(\delta_{bc}^2) \,\right] \,,
\eqe
where
\eqb
 T_N = \frac{1}{6 \pi M}\,
       \left[\, 1 - 4 \left(\frac{\delta_w}{\delta_{bc}}\right)^2 \,\right]^{-1/2}\,
       \left[\, 1 + O(\delta_w) + O(\delta_{bc}^2) + O(\delta_w\,\delta_{bc}^2) \,\right] \,.
\eqe
\seqe
This means that, in the near Nariai case, the temperatures of our two thermal equilibrium systems are equal up to the leading term. 
Hence, at the leading term approximation, the total system composed of two horizons is in a thermal equilibrium state in the near Nariai case. 
Then, as discussed in the first paragraph in Sec.\ref{sec:sd.1}, we expect that the entropy-area law holds at the leading term approximation in the near Nariai case. 
To see it, let us show the free energies~\eref{eq:beh.Fb} and~\eref{eq:ceh.Fc} in the near Nariai case,
\seqb
\eqab
 F_b &=& 3 M - \frac{3 M}{2}
             \left[\, 1 - 4 \left(\frac{\delta_w}{\delta_{bc}}\right)^2 \,\right]^{-1/2}
      + O(\delta_{bc}) + O(\delta_w) + O(\delta_{bc}\,\delta_w) \\
 F_c &=& F_b - 3 M \,.
\eqae
\seqe
Here one may think that the difference $3M = F_b - F_c$ results in the difference between entropies of BEH and CEH. 
But the definition of entropy, $S \defeq -\partial F(T,A,X)/\partial T$, is important. 
Up to the leading term, the system size is $A = 4 \pi (3 M)^2$ and $M$ is fixed in calculating the entropy. 
Therefore the difference $3M$ does not mean the difference between horizon entropies. 
The entropy of BEH is equal to that of CEH at the leading term approximation in the near Nariai case. 
However, unfortunately, we can not check if the entropy-area law recovers at the leading term approximation, because the state variables $X_b$ and $X_c$ are not specified and the partial derivative $-\partial F/\partial A$ can not be calculated. 
At present, we simply expect that the entropy-area law holds at the leading term approximation in the near Nariai limit.


\end{document}